\documentclass[11pt]{article}
\newcommand{\bq}{\begin{quotation}\noindent}
\newcommand{\eq}{\end{quotation}}
\newcommand{\be}{\begin{equation}}
\newcommand{\ee}{\end{equation}}
\newcommand{\bea}{\begin{eqnarray}}
\newcommand{\eea}{\end{eqnarray}}
\newcommand{\bc}{\begin{center}}
\newcommand{\ec}{\end{center}}
\def\tr{{\rm tr}}

\usepackage{graphicx}
\usepackage{indentfirst}

\begin{document}

\title{Quantum Mechanics as Quantum Information \\ (and only a little more)}

\author{Christopher A. Fuchs \smallskip
\\
\it Computing Science Research Center
\\
\it Bell Labs, Lucent Technologies
\\
\it Room 2C-420, 600--700 Mountain Ave.
\\
\it Murray Hill, New Jersey 07974, USA}

\date{}

\maketitle

\bigskip

\raggedbottom

\begin{abstract}
In this paper, I try once again to cause some good-natured trouble.
The issue remains, when will we ever stop burdening the taxpayer with
conferences devoted to the quantum foundations? The suspicion is
expressed that no end will be in sight until a means is found to
reduce quantum theory to two or three statements of crisp physical
(rather than abstract, axiomatic) significance. In this regard, no
tool appears better calibrated for a direct assault than quantum
information theory. Far from a strained application of the latest
fad to a time-honored problem, this method holds promise precisely
because a large part---{\it but not all}---of the structure of
quantum theory has always concerned information.   It is just that
the physics community needs reminding.

This paper, though taking {\tt quant-ph/0106166} as its core,
corrects one mistake and offers several observations beyond the
previous version. In particular, I identify one element of quantum
mechanics that I would {\it not\/} label a subjective term in the
theory---it is the integer parameter $D$ traditionally ascribed to a
quantum system via its Hilbert-space dimension.
\end{abstract}
\bigskip

\section{Introduction\,\protect\footnote{This paper, though
substantially longer, should be viewed as a continuation and
amendment to Ref.~\protect\cite{Fuchs01b}. Details of the changes can
be found in the Appendix to the present paper, Section 11.
Substantial further arguments defending a transition from the
``objective Bayesian'' stance implicit in
Ref.~\protect\cite{Fuchs01b} to the ``subjective Bayesian'' stance
implicit here can be found in Ref.~\protect\cite{Fuchs02a}.}}

Quantum theory as a weather-sturdy structure has been with us for 75
years now. Yet, there is a sense in which the struggle for its
construction remains. I say this because one can check that not a
year has gone by in the last 30 when there was not a meeting or
conference devoted to some aspect of the quantum foundations. Our
meeting in V\"axj\"o, ``Quantum Theory:\ Reconsideration of
Foundations,'' is only one in a long, dysfunctional line.

But how did this come about?  What is the cause of this
year-after-year sacrifice to the ``great mystery?'' Whatever it is,
it cannot be for want of a self-ordained solution: Go to any
meeting, and it is like being in a holy city in great tumult. You
will find all the religions with all their priests pitted in holy
war---the Bohmians\,\cite{Cushing96}, the Consistent
Historians\,\cite{Griffiths99}, the
Transactionalists\,\cite{Cramer88}, the Spontaneous
Collapseans\,\cite{Ghirardi91}, the Einselectionists\,\cite{Zurek98},
the Contextual Objectivists\,\cite{Grangier00}, the outright
Everettics\,\cite{Deutsch97,Vaidman02}, and many more beyond that.
They all declare to see the light, the ultimate light. Each tells us
that if we will accept their solution as our savior, then we too
will see the light.

\footnotesize
\begin{center}
\begin{tabular}{|l||l|}
\hline
\multicolumn{2}{|c|}{} \\
\multicolumn{2}{|c|}{\small \bf A \underline{Fraction} of the Quantum
Foundations Meetings since 1972} \\
\multicolumn{2}{|c|}{} \\
\hline\hline
& \\
1972 & The Development of the Physicist's Conception of Nature,
Trieste, Italy
\\
1973 & Foundations of Quantum Mechanics and Ordered Linear Spaces, \\
& \hspace{.05in} Marbourg, Germany
\\
1974 & Quantum Mechanics, a Half Century Later, Strasbourg, Germany
\\
1975 & Foundational Problems in the Special Sciences, London, Canada
\\
1976 & International Symposium on Fifty Years of the Schr\"odinger
Equation, \\
& \hspace{.05in} Vienna, Austria
\\
1977 & International School of Physics ``Enrico Fermi'',
Course LXXII: \\
& \hspace{.05in} Problems in the Foundations of Physics,
Varenna, Italy
\\
1978 & Stanford Seminar on the Foundations of Quantum Mechanics,
Stanford, USA
\\
1979 & Interpretations and Foundations of Quantum Theory, Marburg,
Germany
\\
1980 & Quantum Theory and the Structures of Time and Space, Tutzing,
Germany
\\
1981 & NATO Advanced Study Institute on Quantum Optics, Experimental
\\
& \hspace{.05in} Gravitation, and Measurement Theory,
Bad Windsheim, Germany
\\
1982 & The Wave-Particle Dualism:\ a Tribute to Louis de Broglie,
Perugia, Italy
\\
1983 & Foundations of Quantum Mechanics in the Light of New
Technology, \\
& \hspace{.05in} Tokyo, Japan
\\
1984 & Fundamental Questions in Quantum Mechanics, Albany, New York
\\
1985 & Symposium on the Foundations of Modern Physics:\ 50 Years of\\
& \hspace{.05in} the Einstein-Podolsky-Rosen Gedankenexperiment,
Joensuu, Finland
\\
1986 & New Techniques and Ideas in Quantum Measurement Theory, New
York, USA
\\
1987 & Symposium on the Foundations of Modern Physics 1987:\ The
Copenhagen\\
& \hspace{.05in} Interpretation 60 Years after the Como Lecture,
Joensuu, Finland
\\
1988 & Bell's Theorem, Quantum Theory, and Conceptions of the
Universe, \\
& \hspace{.05in} Washington, DC, USA
\\
1989 & Sixty-two Years of Uncertainty: Historical, Philosophical
and \\
& \hspace{.05in} Physical Inquiries into the Foundations of Quantum
Mechanics, Erice, Italy
\\
1990 & Symposium on the Foundations of Modern Physics 1990:\ Quantum
Theory of \\
& \hspace{.05in} Measurement and Related Philosophical Problems,
Joensuu, Finland
\\
1991 & Bell's Theorem and the Foundations of Modern Physics, Cesena,
Italy
\\
1992 & Symposia on the Foundations of Modern Physics 1992:\ The
Copenhagen \\
& \hspace{.05in} Interpretation and Wolfgang Pauli, Helsinki, Finland
\\
1993 & International Symposium on Fundamental Problems in Quantum
Physics, \\
& \hspace{.05in} Oviedo, Spain
\\
1994 & Fundamental Problems in Quantum Theory, Baltimore, USA
\\
1995 & The Dilemma of Einstein, Podolsky and Rosen, 60 Years Later,
Haifa, Israel
\\
1996 & 2nd International Symposium on Fundamental Problems in Quantum
Physics, \\
& \hspace{.05in} Oviedo, Spain
\\
1997 & Sixth UK Conference on Conceptual and Mathematical Foundations
of \\
& \hspace{.05in} Modern Physics, Hull, England
\\
1998 & Mysteries, Puzzles, and Paradoxes in Quantum Mechanics, Garda
Lake, Italy
\\
1999 & 2nd Workshop on Fundamental Problems in Quantum Theory,
Baltimore, USA
\\
2000 & NATO Advanced Research Workshop on Decoherence and its
Implications \\
& \hspace{.05in} in Quantum Computation and Information Transfer,
Mykonos, Greece
\\
2001 & Quantum Theory:\ Reconsideration of Foundations, V\"axj\"o,
Sweden\\
&\\
\hline
\end{tabular}
\end{center}
\normalsize\smallskip

But there has to be something wrong with this!  If any of these
priests had truly shown the light, there simply would not be the
year-after-year conference.  The verdict seems clear enough:  If
we---i.e., the set of people who might be reading this paper---really
care about quantum foundations, then it behooves us as a community
to ask why these meetings are happening and find a way to put a stop
to them.

My view of the problem is this. Despite the accusations of
incompleteness, nonsensicality, irrelevance, and surreality one
often sees one religion making against the other, I see little to no
difference in {\it any\/} of their canons.  They all look equally
detached from the world of quantum practice to me. For, though each
seems to want a firm reality within the theory---i.e., a single God
they can point to and declare, ``There, that term is what is real in
the universe even when there are no physicists about''---none have
worked very hard to get out of the Platonic realm of pure
mathematics to find it.

What I mean by this deliberately provocative statement is that in
spite of the differences in what the churches label\footnote{Or {\it
add\/} to the theory, as the case may be.} to be ``real'' in quantum
theory,\footnote{Very briefly, a {\it cartoon\/} of some of the
positions might be as follows.  For the Bohmians, ``reality'' is
captured by supplementing the state vector with an actual trajectory
in coordinate space.  For the Everettics, it is the universal wave
function and the universe's Hamiltonian. (Depending upon the
persuasion, though, these two entities are sometimes supplemented
with the terms in various Schmidt decompositions of the universal
state vector with respect to various preconceived tensor-product
structures.) For the Spontaneous Collapsians it is again the state
vector---though now for the individual system---but Hamiltonian
dynamics is supplemented with an objective collapse mechanism. For
the Consistent Historians ``reality'' is captured with respect to an
initial quantum state and a Hamiltonian by the addition of a set of
preferred positive-operator valued measures (POVMs)---they call them
consistent sets of histories---along with a truth-value assignment
within each of those sets.} they nonetheless all proceed from the
same {\it abstract\/} starting point---the standard textbook
accounts of the {\it axioms\/} of quantum theory.\footnote{To be
fair, they do, each in their own way, contribute minor modifications
to the {\it meanings\/} of a few {\it words\/} in the axioms.  But
that is essentially where the effort
stops.}

\begin{center}
\begin{tabular}{|ll|}
\hline
& \\
\multicolumn{2}{|c|}{\bf The Canon for Most of the Quantum %
                         Churches:} \\
\multicolumn{2}{|c|}{\it \rm The Axioms (plain and simple)} \\
& \\
\hline\hline
& \\
\hspace{.05in} {\bf 1.} & For every system, there is a complex
Hilbert space
$\cal H$.  \\
& \\
\hspace{.05in} {\bf 2.} & States of the system correspond to
projection operators onto $\cal H$. \hspace{.05in} \\
& \\
\hspace{.05in} {\bf 3.} & Those things that are observable {\it
somehow\/}
           correspond to the \\
         & eigenprojectors 
           of Hermitian operators. \\
& \\
\hspace{.05in} {\bf 4.} & Isolated systems evolve according to the
Schr\"odinger  equation. \\
\multicolumn{2}{|c|}{\bf \vdots} \\
& \\
\hline
\end{tabular}
\end{center}

``But what nonsense is this,'' you must be asking.  ``Where else
could they start?''  The main issue is this, and no one has said it
more clearly than Carlo Rovelli\,\cite{Rovelli96}. Where present-day
quantum-foundation studies have stagnated in the stream of history
is not so unlike where the physics of length contraction and time
dilation stood before Einstein's 1905 paper on special relativity.

The Lorentz transformations have the name they do, rather than, say,
the Einstein transformations, for good reason: Lorentz had published
some of them as early as 1895. Indeed one could say that most of the
empirical predictions of special relativity were in place well
before Einstein came onto the scene. But that was of little
consolation to the pre-Einsteinian physics community striving so
hard to make sense of electromagnetic phenomena and the luminiferous
ether. Precisely because the {\it only\/} justification for the
Lorentz transformations appeared to be their {\it empirical
adequacy}, they remained a mystery to be conquered. More
particularly, this was a mystery that heaping further {\it ad hoc\/}
(mathematical) structure onto could not possibly solve.

What was being begged for in the years between 1895 and 1905 was an
understanding of the {\it origin\/} of that abstract, mathematical
structure---some simple, crisp {\it physical\/} statements with
respect to which the necessity of the mathematics would be
indisputable. Einstein supplied that and became one of the greatest
physicists of all time.  He reduced the mysterious structure of the
Lorentz transformations to two simple statements expressible in
common language:
\begin{verse}
1) the speed of light in empty space is independent of the speed of
its source, \\
2) physics should appear the same in all inertial reference frames.
\end{verse}
The deep significance of this for the quantum problem should stand up
and speak overpoweringly to anyone who admires these principles.

Einstein's move effectively stopped all further debate on the origins
of the Lorentz transformations.  Outside of the time of the Nazi
regime in Germany\,\cite{Clark71}, I suspect there have been less
than a handful of conferences devoted to ``interpreting'' them. Most
importantly, with the supreme simplicity of Einstein's principles,
physics became ready for ``the next step.'' Is it possible to
imagine that any mind---even Einstein's---could have made the leap
to general relativity directly from the original, abstract structure
of the Lorentz transformations?  A structure that was only
empirically adequate?  I would say no. Indeed, one can dream of the
wonders we will find in pursuing the same strategy of simplification
for the quantum foundations.\smallskip

\begin{center}
\begin{tabular}{|c||c|} \hline
& \\
\hspace{.05in} \bf Symbolically, where we are: \hspace{.05in} &
\hspace{.05in} \bf Where we need to be: \hspace{.05in} \\
& \\
\hline \hline
& \\
& \\
$\displaystyle x^\prime= \frac{x - v t}{\,\sqrt{1-v^2/c^2}\,}$
& $\matrix{\mbox{Speed of light}\cr\mbox{is constant.}}
$ \\
& \\
& \\
$\displaystyle t^\prime= \frac{t - v x/c^2}{\,\sqrt{1-v^2/c^2}\,}$
& $\matrix{\mbox{Physics is the same}\cr\mbox{in all inertial frames.}}
$ \\
& \\
\hline
\end{tabular}
\end{center}
\smallskip

The task is not to make sense of the quantum axioms by heaping more
structure, more definitions, more science-fiction imagery on top of
them, but to throw them away wholesale and start afresh.  We should
be relentless in asking ourselves:  From what deep {\it physical\/}
principles might we {\it derive\/} this exquisite mathematical
structure?  Those principles should be crisp; they should be
compelling. They should stir the soul. When I was in junior high
school, I sat down with Martin Gardner's book {\sl Relativity for
the Million}\,\cite{Gardner62} and came away with an understanding of
the subject that sustains me today:  The concepts were strange, but
they were clear enough that I could get a grasp on them knowing
little more mathematics than simple arithmetic. One should expect no
less for a proper foundation to quantum theory. Until we can explain
quantum theory's {\it essence\/} to a junior-high-school or
high-school student and have them walk away with a deep, lasting
memory, we will have not understood a thing about the quantum
foundations.

So, throw the existing axioms of quantum mechanics away and start
afresh! But how to proceed? I myself see no alternative but to
contemplate deep and hard the tasks, the techniques, and the
implications of quantum information theory. The reason is simple,
and I think inescapable.  Quantum mechanics has always been about
information.  It is just that the physics community has somehow
forgotten this.

\begin{center}
\begin{tabular}{|l||l|}
\hline
\multicolumn{2}{|c|}{} \\
\multicolumn{2}{|c|}{\small \bf Quantum Mechanics:} \\
\multicolumn{2}{|c|}{\small \it The Axioms and Our Imperative!} \\
\multicolumn{2}{|c|}{} \\
\hline\hline
& \\
\hspace{.05in} \small States correspond to density &
   \small {\it Give an information theoretic} \hspace{.05in} \\
\hspace{.05in}  \small operators $\rho$ over a Hilbert space $\cal
H$.   &
   \hspace{.1in} \small {\it reason if possible!} \\
& \\
\hspace{.05in} \small Measurements correspond to positive & \\
\hspace{.05in} \small operator-valued measures (POVMs) &
   \small {\it Give an information theoretic}\\
\hspace{.05in}   \small $\{E_d\}$ on $\cal H$.   &
   \hspace{.1in} \small {\it reason if possible!} \\
& \\
\hspace{.05in} \small $\cal H$ is a complex vector space, & \\
\hspace{.05in} \small not a real vector space, not a &
   \small {\it Give an information theoretic}\\
\hspace{.05in} \small quaternionic module.  &
   \hspace{.1in} \small {\it reason if possible!} \\
& \\
\hspace{.05in} \small Systems combine according to the tensor & \\
\hspace{.05in} \small product of their separate vector  &
   \small {\it Give an information theoretic}\\
\hspace{.05in} \small spaces, ${\cal H}_{\rm\scriptscriptstyle AB}=
   {\cal H}_{\rm\scriptscriptstyle A}\otimes
   {\cal H}_{\rm\scriptscriptstyle B}$. &
   \hspace{.1in} \small {\it reason if possible!} \\
& \\
\hspace{.05in} \small Between measurements, states evolve & \\
\hspace{.05in} \small according to trace-preserving completely &
   \small {\it Give an information theoretic}\\
\hspace{.05in} \small positive linear maps. &
   \hspace{.1in} \small {\it reason if possible!} \\
& \\
\hspace{.05in} \small By way of measurement, states evolve & \\
\hspace{.05in} \small (up to normalization) via outcome- &
   \small {\it Give an information theoretic}\\
\hspace{.05in} \small dependent completely positive linear maps.
\hspace{.01in} &
   \hspace{.1in} \small {\it reason if possible!} \\
& \\
\hspace{.05in} \small Probabilities for the outcomes & \\
\hspace{.05in} \small of a measurement obey the Born rule &
   \small {\it Give an information theoretic}\\
\hspace{.05in} \small for POVMs ${\rm tr}(\rho E_d)$. &
   \hspace{.1in} \small {\it reason if possible!} \\
& \\
\hline\hline
\multicolumn{2}{|c|}{} \\
\multicolumn{2}{|l|}{\small The distillate that remains---the
piece of quantum theory with no information} \\
\multicolumn{2}{|l|}{\small theoretic significance---will be
our first unadorned glimpse of ``quantum reality.''} \\
\multicolumn{2}{|l|}{\small Far from being
the end of the journey, placing this conception of nature in open} \\
\multicolumn{2}{|l|}{\small
view will be the start of a great adventure.} \\
\multicolumn{2}{|c|}{} \\
\hline
\end{tabular}
\end{center}
\smallskip

This, I see as the line of attack we should pursue with relentless
consistency:  The quantum system represents something real and
independent of us; the quantum state represents a collection of
subjective degrees of belief about {\it something\/} to do with that
system (even if only in connection with our experimental kicks to
it).\footnote{``But physicists are, at bottom, a naive breed,
forever trying to come to terms with the `world out there' by
methods which, however imaginative and refined, involve in essence
the same element of contact as a well-placed kick.'' --- B.~S.
DeWitt and R.~N. Graham\,\cite{DeWitt71}} The structure called
quantum mechanics is about the interplay of these two things---the
subjective and the objective.  The task before us is to separate the
wheat from the chaff.  If the quantum state represents subjective
information, then how much of its mathematical support structure
might be of that same character?  Some of it, maybe most of it, but
surely not all of it.

Our foremost task should be to go to each and every axiom of quantum
theory and give it an information theoretic justification if we can.
Only when we are finished picking off all the terms (or combinations
of terms) that can be interpreted as subjective information will we
be in a position to make real progress in quantum foundations.  The
raw distillate left behind---miniscule though it may be with respect
to the full-blown theory---will be our first glimpse of what quantum
mechanics is trying to tell us about nature itself.

Let me try to give a better way to think about this by making use of
Einstein again. What might have been his greatest achievement in
building general relativity? I would say it was in his recognizing
that the ``gravitational field'' one feels in an accelerating
elevator is a coordinate effect. That is, the ``field'' in that case
is something induced purely with respect to the description of an
observer. In this light, the program of trying to develop general
relativity boiled down to recognizing all the things within
gravitational and motional phenomena that should be viewed as
consequences of our coordinate choices.  It was in identifying all
the things that are ``numerically additional''\,\cite{James67} to
the observer-free situation---i.e., those things that come about
purely by bringing the observer (scientific agent, coordinate
system, etc.)\ back into the picture.

This was a true breakthrough.  For in weeding out all the things that
can be interpreted as coordinate effects, the fruit left behind
finally becomes clear to sight: It is the Riemannian manifold we call
spacetime---a mathematical object, the study of which one can hope
will tell us something about nature itself, not merely about the
observer in nature.

The dream I see for quantum mechanics is just this. Weed out all the
terms that have to do with gambling commitments, information,
knowledge, and belief, and what is left behind will play the role of
Einstein's manifold. That is our goal.  When we find it, it may be
little more than a miniscule part of quantum {\it theory}. But being
a clear window into nature, we may start to see sights through it we
could hardly imagine before.\footnote{I should point out to the
reader that in opposition to the picture of general relativity, where
reintroducing the coordinate system---i.e., reintroducing the
observer---changes nothing about the manifold (it only tells us what
kind of sensations the observer will pick up), I do not suspect the
same for the quantum world. Here I suspect that reintroducing the
observer will be more like introducing matter into pure spacetime,
rather than simply gridding it off with a coordinate system.
``Matter tells spacetime how to curve (when matter is there), and
spacetime tells matter how to move (when matter is
there).''\,\cite{Misner73} Observers, scientific agents, a necessary
part of reality?  No.  But do they tend to change things once they
are on the scene.  Yes.  If quantum mechanics can tell us something
truly deep about nature, I think it is this.}

\section{Summary}
\begin{flushright}
\parbox{4.0in}{\footnotesize
\bq
I say to the House as I said to ministers who have joined this
government, I have nothing to offer but blood, toil, tears, and
sweat. We have before us an ordeal of the most grievous kind. We
have before us many, many months of struggle and suffering. You ask,
what is our policy? I say it is to wage war. War with all our might
and with all the strength God has given us. You ask, what is our
aim?  I can answer in one word. It is \smallskip victory.
\\
\hspace*{\fill} --- {\it Winston Churchill}, 1940, abridged
\eq
}
\end{flushright}

This paper is about taking the imperative in the Introduction
seriously, though it contributes only a small amount to the labor it
asks. Just as in the founding of quantum mechanics, this is not
something that will spring forth from a lone mind in the shelter of
a medieval college.\footnote{If you want to know what this means,
ask me over a beer sometime.} It is a task for a community with
diverse but productive points of view. The quantum information
community is nothing if not that.\footnote{There have been other
soundings of the idea that information and computation theory can
tell us something deep about the foundations of quantum mechanics.
See Refs.~\cite{Zeilinger01}, \cite{Devetak02}, \cite{Mermin00}, and
in particular Ref.~\cite{Barnum02}.} ``Philosophy is too important to
be left to the philosophers,'' John Archibald Wheeler once said.
Likewise, I am apt to say for the quantum foundations.

The structure of the remainder of the paper is as follows.  In
Section 3 ``{\bf Why Information?},'' I reiterate the cleanest
argument I know of that the quantum state is solely an expression of
subjective information---the information one has about a quantum
system. It has no objective reality in and of itself.\footnote{In
the previous version of this paper, {\tt quant-ph/0106166}, I
variously called quantum states ``information'' and ``states of
knowledge'' and did not emphasize so much the ``radical'' Bayesian
idea that the probability one {\it ascribes\/} to a phenomenon
amounts to {\it nothing\/} more than the gambling commitments one is
willing to make with regard to that phenomenon. To the ``radical''
Bayesian, probabilities are subjective all the way to the bone.  In
this paper, I start the long process of trying to turn my earlier
de-emphasis around (even though it is somewhat dangerous to attempt
this in a manuscript that is little more than a modification of an
already completed paper). In particular, because of the objective
overtones of the word ``knowledge''---i.e., that a particular piece
of knowledge is either ``right'' or ``wrong''---I try to steer clear
from the term as much as possible in the present version. The
conception working in the background of this paper is that there is
simply no such thing as a ``right and true'' quantum state.  In all
cases, a quantum state is specifically and only a mathematical symbol
for capturing a set of beliefs or gambling commitments.  Thus I now
variously call quantum states ``beliefs,'' ``states of belief,''
``information'' (though, by this I mean ``information'' in a more
subjective sense than is becoming common in the quantum information
community), ``judgments,'' ``opinions,'' and ``gambling
commitments.'' Believe me, I already understand well the number of
jaws that will drop from the adoption of this terminology. However,
if the reader finds that this gives him a sense of butterflies in
the stomach---or fears that I will become a
solipsist\,\cite{GrangierPrivate} or a crystal-toting New Age
practitioner of homeopathic medicine\,\cite{BennettClub}---I hope he
will keep in mind that this attempt to be absolutely frank about the
subjectivity of \underline{\bf some} of the terms in quantum theory
is part of a larger program to delimit the terms that {\it can be\/}
interpreted as objective in a fruitful way.\label{obfuscatoid}} The
argument is then refined by considering the phenomenon of quantum
teleportation\,\cite{Bennett93}.

In Section 4 ``{\bf Information About What?},'' I tackle that very
question\,\cite{Bub00} head-on.  The answer is ``the potential
consequences of our experimental interventions into nature.'' Once
freed from the notion that quantum measurement ought to be about
revealing traces of some preexisting property\,\cite{Bub97} (or
beable\,\cite{Bell87}), one finds no particular reason to take the
standard account of measurement (in terms of complete sets of
orthogonal projection operators) as a basic notion. Indeed quantum
information theory, with its emphasis on the utility of generalized
measurements or positive operator-valued measures
(POVMs)\,\cite{Nielsen00}, suggests one should take those entities
as the basic notion instead.  The productivity of this point of view
is demonstrated by the enticingly simple Gleason-like derivation of
the quantum probability rule recently found by Paul
Busch\,\cite{Busch99} and, independently, by Joseph Renes and
collaborators\,\cite{Renes00}. Contrary to Gleason's original
theorem\,\cite{Gleason57}, this theorem works just as well for
two-dimensional Hilbert spaces, and even for Hilbert spaces over the
field of rational numbers. In Section 4.1, I give a strengthened
argument for the noncontextuality assumption in this theorem. In
Section 4.2, ``Le Bureau International des Poids et Mesures \`a
Paris,'' I start the process of defining what it means---from the
Bayesian point of view---to accept quantum mechanics as a theory.
This leads to the notion of fixing a fiducial or standard quantum
measurement for defining the very meaning of a quantum state.

In Section 5 ``{\bf Wither Entanglement?},'' I ask whether
entanglement is all it is touted to be as far as quantum foundations
are concerned.  That is, is entanglement really as Schr\"odinger
said, ``{\it the\/} characteristic trait of quantum mechanics, the
one that enforces its entire departure from classical lines of
thought?''  To combat this, I give a simple derivation of the
tensor-product rule for combining Hilbert spaces of individual
systems which takes the structure of {\it localized\/} quantum
measurements as its starting point. In particular, the derivation
makes use of Gleason-like considerations in the presence of classical
communication. With the tensor-product structure established, the
very notion of entanglement follows in step.   This shows how
entanglement, just like the standard probability rule, is secondary
to the structure of quantum measurements. Moreover, ``locality'' is
built in at the outset; there is simply nothing mysterious and
nonlocal about entanglement.

In Section 6 ``{\bf Whither Bayes Rule?},'' I ask why one should
expect the rule for updating quantum state assignments upon the
completion of a measurement to take the form it actually does. Along
the way, I give a simple derivation that one's information always
increases on average for {\it any\/} quantum mechanical measurement
that does not itself discard information.  (Despite the appearance
otherwise, this is not a tautology!) Most importantly, the proof
technique used for showing the theorem indicates an extremely strong
analogy between quantum collapse and Bayes' rule in classical
probability theory: Up to an overall unitary ``readjustment'' of
one's final probabilistic beliefs---the readjustment takes into
account one's initial state for the system as well as one's
description of the measurement interaction---quantum collapse is {\it
precisely\/} Bayesian conditionalization. This in turn gives more
impetus for the assumptions behind the Gleason-like theorems of the
previous two sections.  In Section 6.1, ``Accepting Quantum
Mechanics,'' I complete the process started in Section 4.2 and
describe quantum measurement in Bayesian terms:  An everyday
measurement is any {\it I-know-not-what\/} that leads to an
application of Bayes rule with respect to one's beliefs about the
standard quantum measurement.

In Section 7, ``{\bf What Else is Information?},'' I argue that, to
the extent that a quantum state is a subjective quantity, so must be
the assignment of a state-change rule $\rho\rightarrow\rho_d$ for
describing what happens to an initial quantum state upon the
completion of a measurement---generally some POVM---whose outcome is
$d$. In fact, the levels of subjectivity for the state and the
state-change rule must be precisely the same for consistency's
sake.  To draw an analogy to Bayesian probability theory, the
initial state $\rho$ plays the role of an a priori probability
distribution $P(h)$ for some hypothesis, the final state $\rho_d$
plays the role of a posterior probability distribution $P(h|d)$, and
the state-change rule $\rho\rightarrow\rho_d$ plays the role of the
``statistical model'' $P(d|h)$ enacting the transition
$P(h)\rightarrow P(h|d)$. To the extent that {\it all\/} Bayesian
probabilities are subjective---even the probabilities $P(d|h)$ of a
statistical model---so is the mapping $\rho\rightarrow\rho_d$.
Specializing to the case that no information is gathered, one finds
that the trace-preserving completely positive maps that describe
quantum time-evolution are themselves nothing more than subjective
judgments.

In Section 8 ``{\bf Intermission},'' I give a slight breather to sum
up what has been trashed and where we are headed.

In Section 9 ``{\bf Unknown Quantum States?},'' I tackle the
conundrum posed by these very words.  Despite the phrase's
ubiquitous use in the quantum information literature, what can an
{\it unknown\/} state be?  A quantum state---from the present point
of view, explicitly someone's information---must always be known by
someone, if it exists at all.  On the other hand, for many an
application in quantum information, it would be quite contrived to
imagine that there is always someone in the background describing
the system being measured or manipulated, and that what we are doing
is grounding the phenomenon with respect to {\it his\/} state of
belief. The solution, at least in the case of quantum-state
tomography\,\cite{Vogel89}, is found through a quantum mechanical
version of de Finetti's classic theorem on ``unknown
probabilities.''  This reports work from Refs.~\cite{Caves01} and
\cite{Schack00}. Maybe one of the most interesting things about the
theorem is that it fails for Hilbert spaces over the field of real
numbers, suggesting that perhaps the whole discipline of quantum
information might not be well defined in that imaginary world.

Finally, in Section 10 ``{\bf The Oyster and the Quantum},'' I flirt
with the most tantalizing question of all:  Why the quantum? There is
no answer here, but I do not discount that we are on the brink of
finding one. In this regard no platform seems firmer for the leap
than the very existence of quantum cryptography and quantum
computing. The world is sensitive to our touch.  It has a kind of
``Zing!''\footnote{Dash, verve, vigor, vim, zip, pep, punch,
pizzazz!} that makes it fly off in ways that were not imaginable
classically. The whole structure of quantum mechanics---{\it
\underline{it is speculated}}---may be nothing more than the optimal
method of reasoning and processing information in the light of such a
fundamental (wonderful) sensitivity.  As a concrete proposal for a
potential mathematical expression of ``Zing!,'' I consider the
integer parameter $D$ traditionally ascribed to a quantum system by
way of its Hilbert-space dimension.

\section{Why Information?}
\begin{flushright}
\parbox{4.0in}{\footnotesize
\bq
Realists can be tough customers indeed---but there is no reason to
be afraid of them.\\
\hspace*{\fill} --- {\it Paul Feyerabend, 1992}\\
\eq
}
\end{flushright}

Einstein was the master of clear thought; I have expressed my opinion
of this with respect to both special and general relativity.  But I
can go further.  I would say he possessed the same great penetrating
power when it came to analyzing the quantum. For even there, he was
immaculately clear and concise in his expression. In particular, he
was the first person to say in absolutely unambiguous terms why the
quantum state should be viewed as information (or, to say the same
thing, as a representation of one's beliefs and gambling commitments,
credible or otherwise).

His argument was simply that a quantum-state assignment for a system
can be forced to go one way or the other by interacting with a part
of the world that should have no causal connection with the system
of interest.  The paradigm here is of course the one well known
through the Einstein, Podolsky, Rosen paper\,\cite{Einstein35}, but
simpler versions of the train of thought had a long pre-history with
Einstein\,\cite{Fine86} himself.

The best was in essence this.  Take two spatially separated systems
$A$ and $B$ prepared in some entangled quantum state
$|\psi^{\scriptscriptstyle AB}\rangle$. By performing the
measurement of one or another of two observables on system $A$
alone, one can {\it immediately\/} write down a new state for system
$B$.  Either the state will be drawn from one set of states
$\{|\phi_i^{\scriptscriptstyle B}\rangle\}$ or another
$\{|\eta_i^{\scriptscriptstyle B}\rangle\}$, depending upon which
observable is measured.\footnote{Generally there need be hardly any
relation between the two sets of states:  only that when the states
are weighted by their probabilities, they mix together to form the
initial density operator for system $B$ alone. For a precise
statement of this freedom, see Ref.~\cite{Hughston93}.} The key point
is that it does not matter how distant the two systems are from each
other, what sort of medium they might be immersed in, or any of the
other fine details of the world.  Einstein concluded that whatever
these things called quantum states {\it be}, they cannot be ``real
states of affairs'' for system $B$ alone.  For, whatever the real,
objective state of affairs at $B$ is, it should not depend upon the
measurements one can make on a causally unconnected system $A$.

Thus one must take it seriously that the new state (either a
$|\phi_i^{\scriptscriptstyle B}\rangle$ or a
$|\eta_i^{\scriptscriptstyle B}\rangle$) represents information
about system $B$.  In making a measurement on $A$, one learns
something about $B$, but that is where the story ends. The state
change cannot be construed to be something more physical than that.
More particularly, the final state itself for $B$ cannot be viewed as
more than a reflection of some tricky combination of one's initial
information and the knowledge gained through the measurement.
Expressed in the language of Einstein, the quantum state cannot be a
``complete'' description of the quantum system.

Here is the way Einstein put it to Michele Besso in a 1952
letter\,\cite{Bernstein91}:
\begin{quotation}
\small What relation is there between the ``state'' (``quantum
state'') described by a function $\psi$ and a real deterministic
situation (that we call the ``real state'')?  Does the quantum state
characterize completely (1) or only incompletely (2) a real state?

One cannot respond unambiguously to this question, because each
measurement represents a real uncontrollable intervention in the
system (Heisenberg). The real state is not therefore something that
is immediately accessible to experience, and its appreciation always
rests hypothetical. (Comparable to the notion of force in classical
mechanics, if one doesn't fix {\it a priori\/} the law of motion.)
Therefore suppositions (1) and (2) are, in principle, both possible.
A decision in favor of one of them can be taken only after an
examination and confrontation of the admissibility of their
consequences.

I reject (1) because it obliges us to admit that there is a rigid
connection between parts of the system separated from each other in
space in an arbitrary way (instantaneous action at a distance, which
doesn't diminish when the distance increases).  Here is the
demonstration:

A system $S_{12}$, with a function $\psi_{12}$, which is known, is
composed of two systems $S_1$, and $S_2$, which are very far from
each other at the instant $t$. If one makes a ``complete''
measurement on $S_1$, which can be done in different ways (according
to whether one measures, for example, the momenta or the
coordinates), depending on the result of the measurement and the
function $\psi_{12}$, one can determine by current
quantum-theoretical methods, the function $\psi_2$ of the second
system. {\it This function can assume different forms}, according to
the {\it procedure\/} of measurement applied to $S_1$.

But this is in contradiction with (1) {\it if one excludes action at
a distance}. Therefore the measurement on $S_1$ has no effect on the
real state $S_2$, and therefore assuming (1) no effect on the quantum
state of $S_2$ described by $\psi_2$.

I am thus forced to pass to the supposition (2) according to which
the real state of a system is only described incompletely by the
function $\psi_{12}$.

If one considers the method of the present quantum theory as being in
principle definitive, that amounts to renouncing a complete
description of real states.  One could justify this renunciation if
one assumes that there is no law for real states---i.e., that their
description would be useless.  Otherwise said, that would mean: laws
don't apply to things, but only to what observation teaches us about
them.  (The laws that relate to the temporal succession of this
partial knowledge are however entirely deterministic.)

Now, I can't accept that.  I think that the statistical character of
the present theory is simply conditioned by the choice of an
incomplete description.
\end{quotation}

There are two issues in this letter that are worth disentangling. 1)
Rejecting the rigid connection of all nature\footnote{The rigid
connection of all nature, on the other hand, is exactly what the
Bohmians and Everettics {\it do\/} embrace, even glorify.  So, I
suspect these words will fall on deaf ears with them.  But similarly
would they fall on deaf ears with the believer who says that God
wills each and every event in the universe and no further
explanation is needed. No point of view should be dismissed out of
hand:  the overriding issue is simply which view will lead to the
most progress, which view has the potential to close the debate,
which view will give the most new phenomena for the physicist to
have fun with? \label{GodandAllah}}---that is to say, admitting that
the very notion of {\it separate systems\/} has any meaning at
all---one is led to the conclusion that a quantum state cannot be a
complete specification of a system.  It must be information, at
least in part.  This point should be placed in contrast to the other
well-known facet of Einstein's thought: namely, 2) an unwillingness
to accept such an ``incompleteness'' as a necessary trait of the
physical world.

It is quite important to recognize that the first issue does not
entail the second.  Einstein had that firmly in mind, but he wanted
more. His reason for going the further step was, I think, well
justified {\it at the time\/}\,\cite{Einstein70}:
\begin{quotation}
\small There exists \ldots\ a simple psychological reason for the fact
that this most nearly obvious interpretation is being shunned.  For
if the statistical quantum theory does not pretend to describe the
individual system (and its development in time) completely, it
appears unavoidable to look elsewhere for a complete description of
the individual system; in doing so it would be clear from the very
beginning that the elements of such a description are not contained
within the conceptual scheme of the statistical quantum theory.  With
this one would admit that, in principle, this scheme could not serve
as the basis of theoretical physics.
\end{quotation}

But the world has seen much in the mean time.  The last seventeen
years have given confirmation after confirmation that the Bell
inequality (and several variations of it) are indeed violated by the
physical world.  The Kochen-Specker no-go theorems have been
meticulously clarified to the point where simple textbook pictures
can be drawn of them\,\cite{Peres93}.  Incompleteness, it seems, is
here to stay:  The theory prescribes that no matter how much we know
about a quantum system---even when we have {\it maximal\/}
information about it\,\footnote{As should be clear from all my
warnings, I am no longer entirely pleased with this terminology.  I
would now, for instance, refer to a pure quantum state as a
``maximally rigid gambling commitment'' or some such thing. See
Ref.~\cite{Fuchs02a}, pages 49--50 and 53--54. However, after trying
to reconstruct this paragraph several times to be in conformity with
my new terminology, I finally decided that a more accurate
representation would break the flow of the section even more than
this footnote!}
---there will always be a statistical residue. There will always be
questions that we can ask of a system for which we cannot predict
the outcomes.  In quantum theory, maximal information is simply not
complete information\,\cite{Caves96}.  But neither can it be
completed.  As Wolfgang Pauli once wrote to Markus
Fierz\,\cite{Pauli54}, ``The well-known `incompleteness' of quantum
mechanics (Einstein) is certainly an existent fact
somehow-somewhere, but certainly cannot be removed by reverting to
classical field physics.''  Nor, I would add, will the mystery of
that ``existent fact'' be removed by attempting to give the quantum
state anything resembling an ontological status.

The complete disconnectedness of the quantum-state change rule from
anything to do with spacetime considerations is telling us something
deep: The quantum state is information. Subjective, incomplete
information. Put in the right mindset, this is {\it not\/} so
intolerable.  It is a statement about our world. There is something
about the world that keeps us from ever getting more information
than can be captured through the formal structure of quantum
mechanics. Einstein had wanted us to look further---to find out how
the incomplete information could be completed---but perhaps the real
question is, ``Why can it {\it not\/} be completed?''

Indeed I think this is one of the deepest questions we can ask and
still hope to answer. But first things first. The more immediate
question for anyone who has come this far---and one that deserves to
be answered forthright---is what is this information symbolized by a
$|\psi\rangle$ actually about? I have hinted that I would not dare
say that it is about some kind of hidden variable (as the Bohmian
might) or even about our place within the universal wavefunction (as
the Everettic might).

Perhaps the best way to build up to an answer is to be true to the
theme of this paper:  quantum foundations in the light of quantum
information.  Let us forage the phenomena of quantum information to
see if we might first refine Einstein's argument.  One need look no
further than to the phenomenon of quantum
teleportation\,\cite{Bennett93}.  Not only can a quantum-state
assignment for a system be forced to go one way or the other by
interacting with another part of the world of no causal significance,
but, for the cost of two bits, one can make that quantum state
assignment anything one wants it to be.

Such an experiment starts out with Alice and Bob sharing a maximally
entangled pair of qubits in the state
\be
|\psi^{\scriptscriptstyle AB}\rangle=
\sqrt{\frac{1}{2}}\,\big(|0\rangle|0\rangle+|1\rangle
|1\rangle\big)\;.
\ee
Bob then goes to any place in the universe he wishes. Alice in her
laboratory prepares another qubit with any state $|\psi\rangle$ that
she ultimately wants to impart onto Bob's system.  She performs a
Bell-basis measurement on the two qubits in her possession.  In the
same vein as Einstein's thought experiment, Bob's system immediately
takes on the character of one of the states $|\psi\rangle$, $\sigma_x
|\psi\rangle$, $\sigma_y |\psi\rangle$, or $\sigma_z |\psi\rangle$.
But that is only insofar as Alice is concerned.\footnote{As far as
Bob is concerned, nothing whatsoever changes about the system in his
possession:  It started in the completely mixed state $\rho =
\frac{1}{2}I$ and remains that way.} Since there is no (reasonable)
causal connection between Alice and Bob, it must be that these
states represent the possibilities for Alice's updated beliefs about
Bob's system.

If now Alice broadcasts the result of her measurement to the world,
Bob may complete the teleportation protocol by performing one of the
four Pauli rotations ($I$, $\sigma_x$, $\sigma_y$, $\sigma_z$) on
his system, conditioning it on the information he receives.  The
result, as far as Alice is concerned, is that Bob's system finally
resides predictably in the state $|\psi\rangle$.\footnote{As far as
Bob is concerned, nothing whatsoever changes about the system in his
possession:  It started in the completely mixed state $\rho =
\frac{1}{2}I$ and remains that way.}\footnote{The repetition in
these footnotes is not a typographical error.}

How can Alice convince herself that such is the case?  Well, if Bob
is willing to reveal his location, she just need walk to his site and
perform the {\bf YES}-{\bf NO} measurement:
$|\psi\rangle\langle\psi|$ vs.\ $I-|\psi\rangle\langle\psi|$.  The
outcome will be a {\bf YES} with probability one for her if all has
gone well in carrying out the protocol.  Thus, for the cost of a
measurement on a causally disconnected system and two bits worth of
causal action on the system of actual interest---i.e., one of the
four Pauli rotations---Alice can sharpen her predictability to
complete certainty for {\it any\/} {\bf YES}-{\bf NO} observable she
wishes.

Roger Penrose argues in his book {\sl The Emperor's New Mind\/}
\,\cite{Penrose89} that when a system ``has'' a state $|\psi\rangle$
there ought to be some property in the system (in and of itself) that
corresponds to its ``$|\psi\rangle$'ness.'' For how else could the
system be prepared to reveal a {\bf YES} in the case that Alice
actually checks it?  Asking this rhetorical question with a
sufficient amount of command is enough to make many a would-be
informationist weak in knees.  But there is a crucial oversight
implicit in its confidence, and we have already caught it in
action.  If Alice fails to reveal her information to anyone else in
the world, there is no one else who can predict the qubit's ultimate
revelation with certainty.  More importantly, there is nothing in
quantum mechanics that gives the qubit the power to stand up and say
{\bf YES} all by itself:  If Alice does not take the time to walk
over to it and interact with it, there is no revelation. There is
only the confidence in Alice's mind that, {\it should\/} she
interact with it, she {\it could\/} predict the
consequence\,\footnote{I adopt this terminology to be similar to
L.~J. Savage's book, Ref.~\cite{Savage54}, Chapter 2, where he
discusses the terms ``the person,'' ``the world,'' ``consequences,''
``acts,'' and ``decisions,'' in the context of rational decision
theory. ``A {\bf consequence} is anything that may happen to the
person,'' Savage writes, where we add ``when he {\bf acts} via the
capacity of a quantum measurement.'' In this paper, I call what
Savage calls ``the person'' the agent, scientific agent, or observer
instead.} of that interaction.

\section{Information About What?}
\begin{flushright}
\parbox{4.1in}{\footnotesize
\bq
I think that the sickliest notion of physics, even if a student gets
it, is that it is `the science of masses, molecules, and the
ether.'  And I think that the healthiest notion, even if a student
does not wholly get it, is that physics is the science of the ways
of taking hold of bodies and pushing them!
\\
\hspace*{\fill} --- {\it W.~S. Franklin, 1903}
\eq
}
\end{flushright}

There are great rewards in being a new parent.  Not least of all is
the opportunity to have a close-up look at a mind in formation. Last
year, I watched my two-year old daughter learn things at a fantastic
rate, and though there were untold lessons for her, there were also
a sprinkling for me.  For instance, I started to see her come to
grips with the idea that there is a world independent of her
desires. What struck me was the contrast between that and the gain of
confidence I also saw grow in her that there are aspects of existence
she {\it could\/} control.  The two go hand in hand.  She pushes on
the world, and sometimes it gives in a way that she has learned to
predict, and sometimes it pushes back in a way she has not foreseen
(and may never be able to). If she could manipulate the world to the
complete desires of her will---I became convinced---there would be
little difference between wake and dream.

The main point is that she learns from her forays into the world. In
my cynical moments, I find myself thinking, ``How can she think that
she's learned anything at all?  She has no theory of measurement.
She leaves measurement completely undefined.  How can she have a
stake to knowledge if she does not have a theory of how she learns?''

Hideo Mabuchi once told me, ``The quantum measurement problem refers
to a set of people.''  And though that is a bit harsh, maybe it also
contains a bit of the truth.  With the physics community making use
of theories that tend to last between 100 and 300 years, we are apt
to forget that scientific views of the world are built from the top
down, not from the bottom up.  The experiment is the basis of all
which we try to describe with science.  But an experiment is an
active intervention into the course of nature on the part of the
experimenter; it is not contemplation of nature from
afar\,\cite{Jammer84}. We set up this or that experiment to see how
nature reacts. It is the conjunction of myriads of such
interventions and their consequences that we record into our data
books.\footnote{But I must stress that I am not so positivistic as
to think that physics should somehow be grounded on a primitive
notion of ``sense impression'' as the philosophers of the Vienna
Circle did.  The interventions and their consequences that an
experimenter records, have no option but to be thoroughly
theory-laden.  It is just that, in a sense, they are by necessity at
least one theory behind.  No one got closer to the salient point than
Heisenberg (in a quote he attributed to Einstein many years after the
fact)\,\cite{Heisenberg71}:
\begin{quote}
It is quite wrong to try founding a theory on observable magnitudes
alone.  In reality the very opposite happens.  It is the theory which
decides what we can observe.  You must appreciate that observation is
a very complicated process.  The phenomenon under observation
produces certain events in our measuring apparatus.  As a result,
further processes take place in the apparatus, which eventually and
by complicated paths produce sense impressions and help us to fix the
effects in our consciousness.  Along this whole path---from the
phenomenon to its fixation in our consciousness---we must be able to
tell how nature functions, must know the natural laws at least in
practical terms, before we can claim to have observed anything at
all.  Only theory, that is, knowledge of natural laws, enables us to
deduce the underlying phenomena from our sense impressions.  When we
claim that we can observe something new, we ought really to be saying
that, although we are about to formulate new natural laws that do not
agree with the old ones, we nevertheless assume that the existing
laws---covering the whole path from the phenomenon to our
consciousness---function in such a way that we can rely upon them and
hence speak of ``observation.''
\end{quote}\vspace{-.15in}}

We tell ourselves that we have learned something new when we can
distill from the data a compact description of all that was seen
and---even more tellingly---when we can dream up further experiments
to corroborate that description. This is the minimal requirement of
science.  If, however, from such a description we can {\it
further\/} distill a model of a free-standing ``reality''
independent of our interventions, then so much the better.  I have
no bone to pick with reality.  It is the most solid thing we can
hope for from a theory.  Classical physics is the ultimate example in
that regard. It gives us a compact description, but it can give much
more if we want it to.

The important thing to realize, however, is that there is no logical
necessity that such a worldview always be obtainable.  If the world
is such that we can never identify a reality---a {\it
free-standing\/} reality---independent of our experimental
interventions, then we must be prepared for that too. That is where
quantum theory in its most minimal and conceptually simplest
dispensation seems to stand\,\cite{Fuchs00}. It is a theory whose
terms refer predominately to our interface with the world.  It is a
theory that cannot go the extra step that classical physics did
without ``writing songs I can't believe, with words that tear and
strain to rhyme''\,\cite{Simon65}. It is a theory not about
observables, not about beables, but about
``dingables.''\footnote{Pronounced ding-ables.} We tap a bell with
our gentle touch and listen for its beautiful ring.

So what are the ways we can intervene on the world?  What are the
ways we can push it and wait for its unpredictable reaction?  The
usual textbook story is that those things that are measurable
correspond to Hermitian operators.  Or perhaps to say it in more
modern language, to each observable there corresponds a set of
orthogonal projection operators $\{\Pi_i\}$ over a complex Hilbert
space ${\cal H}_{\scriptscriptstyle\rm D}$ that form a complete
resolution of the identity,
\be
\sum_i \Pi_i = I\;.
\ee
The index $i$ labels the potential outcomes of the measurement (or
{\it intervention\/}, to slip back into the language promoted above).
When an observer possesses the information $\rho$---captured most
generally by a mixed-state density operator---quantum mechanics
dictates that he can expect the various outcomes with a probability
\be
P(i)=\tr (\rho \Pi_i)\;.
\ee

The best justification for this probability rule comes by way of
Andrew Gleason's amazing 1957 theorem\,\cite{Gleason57}.  For, it
states that the standard rule is the {\it only\/} rule that
satisfies a very simple kind of noncontextuality for measurement
outcomes\,\cite{Barnum00}. In particular, if one contemplates
measuring two distinct observables $\{\Pi_i\}$ and $\{\Gamma_i\}$
which happen to share a single projector $\Pi_k$, then the
probability of outcome $k$ is independent of which observable it is
associated with.  More formally, the statement is this.  Let ${\cal
P}_{\scriptscriptstyle\rm D}$ be the set of projectors associated
with a (real or complex) Hilbert space ${\cal
H}_{\scriptscriptstyle\rm D}$ for $D\ge3$, and let $f:{\cal
P}_{\scriptscriptstyle\rm D}\longrightarrow[0,1]$ be such that
\be
\sum_i f(\Pi_i) = 1
\ee
whenever a set of projectors $\{\Pi_i\}$ forms an observable. The
theorem concludes that there exists a density operator $\rho$ such
that
\be
f(\Pi)=\tr (\rho \Pi)\;.
\ee
In fact, in a single blow, Gleason's theorem derives not only the
probability rule, but also the state-space structure for quantum
mechanical states (i.e., that it corresponds to the convex set of
density operators).

In itself this is no small feat, but the thing that makes the theorem
an ``amazing'' theorem is the sheer difficulty required to prove
it\,\cite{Cooke81}. Note that no restrictions have been placed upon
the function $f$ beyond the ones mentioned above. There is no
assumption that it need be differentiable, nor that it even need be
continuous.  All of that, and linearity too, comes from the
structure of the observables---i.e., that they are complete sets of
orthogonal projectors onto a linear vector space.

Nonetheless, one should ask:  Does this theorem really give {\it the
physicist\/} a clearer vision of where the probability rule comes
from? Astounding feats of mathematics are one thing; insight into
physics is another. The two are often at opposite ends of the
spectrum. As fortunes turn, a unifying strand can be drawn by
viewing quantum foundations in the light of quantum information.

The place to start is to drop the fixation that the basic set of
observables in quantum mechanics are complete sets of orthogonal
projectors.  In quantum information theory it has been found to be
extremely convenient to expand the notion of measurement to also
include general positive operator-valued measures
(POVMs)\,\cite{Peres93,Kraus83}.  In other words, in place of the
usual textbook notion of measurement, {\it any\/} set $\{E_d\}$ of
positive-semidefinite operators on ${\cal H}_{\scriptscriptstyle\rm
D}$ that forms a resolution of the identity, i.e., that satisfies
\begin{equation}
\langle\psi|E_d|\psi\rangle\ge0\,,\quad\mbox{for all
$|\psi\rangle\in{\cal H}_{\scriptscriptstyle\rm D}$}
\label{Hank}
\end{equation}
and
\begin{equation}
\sum_d E_d = I\;,
\label{Hannibal}
\end{equation}
counts as a measurement. The outcomes of the measurement are
identified with the indices $d$, and the probabilities of the
outcomes are computed according to a generalized Born rule,
\begin{equation}
P(d)=\tr\big(\rho E_d\big) \;.
\label{ChickenAndGravy}
\end{equation}
The set $\{E_d\}$ is called a POVM, and the operators $E_d$ are
called POVM elements. (In the nonstandard language promoted earlier,
the set $\{E_d\}$ signifies an intervention into nature, while the
individual $E_d$ represent the potential consequences of that
intervention.) Unlike standard measurements, there is no limitation
on the number of values the index $d$ can take. Moreover, the $E_d$
may be of any rank, and there is no requirement that they be
mutually orthogonal.

The way this expansion of the notion of measurement is {\it
usually\/} justified is that any POVM can be represented formally as
a standard measurement on an ancillary system that has interacted in
the past with the system of actual interest. Indeed, suppose the
system and ancilla are initially described by the density operators
$\rho_{\scriptscriptstyle\rm S}$ and $\rho_{\scriptscriptstyle\rm A}$
respectively. The conjunction of the two systems is then described
by the initial quantum state
\begin{equation}
\rho_{\scriptscriptstyle\rm SA}=\rho_{\scriptscriptstyle\rm
S}\otimes\rho_{\scriptscriptstyle\rm A}\;.
\label{MushMush}
\end{equation}
An interaction between the systems via some unitary time evolution
leads to a new state
\begin{equation}
\rho_{\scriptscriptstyle\rm SA}\;\longrightarrow\;
U\rho_{\scriptscriptstyle\rm SA} U^\dagger\;.
\label{Aluminium}
\end{equation}
Now, imagine a standard measurement on the ancilla.  It is described
on the total Hilbert space via a set of orthogonal projection
operators $\{I\otimes\Pi_d\}$. An outcome $d$ will be found, by the
standard Born rule, with probability
\begin{equation}
P(d)={\rm tr}\!\left( U(\rho_{\scriptscriptstyle\rm
S}\otimes\rho_{\scriptscriptstyle\rm A})
U^\dagger(I\otimes\Pi_d)\right)\;.
\label{RibTie}
\end{equation}
The number of outcomes in this seemingly indirect notion of
measurement is limited only by the dimensionality of the ancilla's
Hilbert space---in principle, there can be arbitrarily many.

As advertised, it turns out that the probability formula above can
be expressed in terms of operators on the system's Hilbert space
alone: This is the origin of the POVM\@. If we let $|s_\alpha\rangle$
and $|a_c\rangle$ be an orthonormal basis for the system and ancilla
respectively, then $|s_\alpha\rangle|a_c\rangle$ will be a basis for
the composite system. Using the cyclic property of the trace in
Eq.~(\ref{RibTie}), we get
\bea
P(d)
&=&
\sum_{\alpha c}\langle s_\alpha|\langle a_c|\Big( (\rho_{\rm
s}\otimes\rho_{\scriptscriptstyle\rm A})  U^\dagger(I\otimes\Pi_d)
U\Big)|s_\alpha \rangle|a_c\rangle
\nonumber\\
&=&
\sum_\alpha\langle s_\alpha|\,\rho_{\scriptscriptstyle\rm
S}\!\left(\sum_c\langle a_c|\Big(
(I\otimes\rho_{\scriptscriptstyle\rm A}) U^\dagger(I \otimes\Pi_d)
U\Big)| a_c\rangle\!\right)\!|s_\alpha\rangle\;. \rule{0mm}{8mm}
\eea
Letting ${\rm tr}_{\scriptscriptstyle\rm A}$ and ${\rm
tr}_{\scriptscriptstyle\rm S}$ denote partial traces over the system
and ancilla, respectively, it follows that
\be
P(d)={\rm tr}_{\scriptscriptstyle\rm S}(\rho_{\scriptscriptstyle\rm
S} E_d)\;,
\ee
where
\be
E_d={\rm tr}_{\scriptscriptstyle\rm A}\!\left(
(I\otimes\rho_{\scriptscriptstyle\rm A}) U(I\otimes\Pi_d)
U^\dagger\right)
\label{oncogene}
\ee
is an operator acting on the Hilbert space of the original system.
This proves half of what is needed, but it is also straightforward
to go in the reverse direction---i.e., to show that for any POVM
$\{E_d\}$, one can pick an ancilla and find operators
$\rho_{\scriptscriptstyle\rm A}$, $U$, and $\Pi_d$ such that
Eq.~(\ref{oncogene}) is true.

Putting this all together, there is a sense in which standard
measurements capture everything that can be said about quantum
measurement theory\,\cite{Kraus83}.  As became clear above, a way to
think about this is that by learning something about the ancillary
system through a standard measurement, one in turn learns something
about the system of real interest. Indirect though it may seem, this
can be a powerful technique, sometimes revealing information that
could not have been revealed otherwise\,\cite{Holevo73}.  A very
simple example is where a sender has only a single qubit available
for the sending one of three potential messages. She therefore has a
need to encode the message in one of three preparations of the
system, even though the system is a two-state system. To recover as
much information as possible, the receiver might (just intuitively)
like to perform a measurement with three distinct outcomes.  If,
however, he were limited to a standard quantum measurement, he would
only be able to obtain two outcomes. This---perhaps
surprisingly---generally degrades his opportunities for recovery.

What I would like to bring up is whether this standard way of
justifying the POVM is the most productive point of view one can
take.  Might any of the mysteries of quantum mechanics be alleviated
by taking the POVM as a basic notion of measurement?  Does the
POVM's utility portend a larger role for it in the foundations of
quantum mechanics?

\begin{center}
\begin{tabular}{|c||c|}
\hline
& \\
\bf Standard & \bf Generalized
\\
\bf Measurements & \bf Measurements
\\
& \\
\hline\hline
& \\
$\{\Pi_i\}$ & $\{E_d\}$
\\
& \\
\hspace{.1in} $\langle\psi|\Pi_i|\psi\rangle\ge0\,,\;\forall
|\psi\rangle$ \hspace{.1in}
& \hspace{.1in} $\langle\psi|E_d|\psi\rangle\ge0\,,\;\forall
|\psi\rangle$ \hspace{.1in}
\\
& \\
$\sum_i \Pi_i = I$ & $\sum_d E_d = I$
\\
& \\
$P(i)=\tr (\rho \Pi_i)$ & $P(d)=\tr (\rho E_d)$
\\
& \\
$\Pi_i\Pi_j=\delta_{ij}\,\Pi_i$ & {\bf ---------}
\\
& \\
\hline
\end{tabular}
\end{center}
\bigskip

I try to make this point dramatic in my lectures by exhibiting a
transparency of the table above.  On the left-hand side there is a
list of various properties for the standard notion of a quantum
measurement. On the right-hand side, there is an almost identical
list of properties for the POVMs.  The only difference between the
two columns is that the right-hand one is {\it missing\/} the
orthonormality condition required of a standard measurement. The
question I ask the audience is this:  Does the addition of that one
extra assumption really make the process of measurement any less
mysterious?  Indeed, I imagine myself teaching quantum mechanics for
the first time and taking a vote with the best audience of all, the
students. ``Which set of postulates for quantum measurement would
you prefer?'' I am quite sure they would respond with a blank stare.
But that is the point! It would make no difference to them, and it
should make no difference to us. The only issue worth debating is
which notion of measurement will allow us to see more deeply into
quantum mechanics.

Therefore let us pose the question that Gleason did, but with POVMs.
In other words, let us suppose that the sum total of ways an
experimenter can intervene on a quantum system corresponds to the
full set of POVMs on its Hilbert space ${\cal
H}_{\scriptscriptstyle\rm D}$.  It is the task of the theory to give
him probabilities for the various consequences of his
interventions.  Concerning those probabilities, let us (in analogy
to Gleason) assume only that whatever the probability for a given
consequence $E_c$ is, it does not depend upon whether $E_c$ is
associated with the POVM $\{E_d\}$ or, instead, any other one
$\{\tilde{E}_d\}$. This means we can assume there exists a function
\be
f:{\cal E}_{\scriptscriptstyle\rm D}\longrightarrow[0,1]\;,
\label{Yaniggle}
\ee
where
\be
{\cal E}_{\scriptscriptstyle\rm
D}=\Big\{E:\,0\le\langle\psi|E|\psi\rangle\le
1\;,\;\forall\;|\psi\rangle\in{\cal H}_{\scriptscriptstyle\rm
D}\Big\}\;,
\ee
such that whenever $\{E_d\}$ forms a POVM,
\be
\sum_d f(E_d) = 1\;.
\ee
(In general, we will call any function satisfying
\be
f(E)\ge0 \qquad \mbox{and} \qquad \sum_d f(E_d) = \mbox{constant}
\ee
a {\it frame function}, in analogy to Gleason's nonnegative frame
functions.  The set ${\cal E}_{\scriptscriptstyle\rm D}$ is often
called the set of {\it effects\/} over ${\cal
H}_{\scriptscriptstyle\rm D}$.)

It will come as no surprise, of course, that a Gleason-like theorem
must hold for the function in Eq.~(\ref{Yaniggle}). Namely, it can
be shown that there must exist a density operator $\rho$ for which
\be
f(E)=\tr (\rho E)\;.
\ee
This was recently shown by Paul Busch\,\cite{Busch99} and,
independently, by Joseph Renes and collaborators\,\cite{Renes00}.
What {\it is\/} surprising however is the utter simplicity of the
proof. Let us exhibit the whole thing right here and now.
\bq
\small
\indent First, consider the case where
${\cal H}_{\scriptscriptstyle\rm D}$ and the operators on it are
defined {\it only\/} over the field of (complex) rational numbers.
It is no problem to see that $f$ is ``linear'' with respect to
positive combinations of operators that never go outside ${\cal
E}_{\rm\scriptscriptstyle D}$.  For consider a three-element POVM
$\{E_1,E_2,E_3\}$. By assumption $f(E_1)+f(E_2)+f(E_3)=1$. However,
we can also group the first two elements in this POVM to obtain a
new POVM, and must therefore have $f(E_1+E_2)+f(E_3)=1$. In other
words, the function $f$ must be additive with respect to a
fine-graining operation:
\be
f(E_1+E_2)= f(E_1)+f(E_2)\;.
\ee
Similarly for any two integers $m$ and $n$,
\be
f(E)=m\, f\!\left(\frac{1}{m}E\right) = n\,
f\!\left(\frac{1}{n}E\right)
\ee
Suppose $\frac{n}{m}\le1$.  Then if we write $E=nG$, this statement
becomes:
\be
f\!\left(\frac{n}{m}G\right)=\frac{n}{m}f(G)\;.
\ee
Thus we immediately have a kind of limited linearity on ${\cal
E}_{\scriptscriptstyle\rm D}$.

One {\it might\/} imagine using this property to cap off the theorem
in the following way.  Clearly the full $D^2$-dimensional vector
space ${\cal O}_{\scriptscriptstyle\rm D}$ of Hermitian operators on
${\cal H}_{\scriptscriptstyle\rm D}$ is spanned by the set ${\cal
E}_{\scriptscriptstyle\rm D}$ since that set contains, among other
things, all the projection operators. Thus, we can write any
operator $E\in{\cal E}_{\scriptscriptstyle\rm D}$ as a linear
combination
\be
E=\sum_{i=1}^{{\scriptscriptstyle D}^2} \alpha_i E_i
\label{Hortense}
\ee
for some fixed operator-basis $\{E_i\}_{i=1}^{{\scriptscriptstyle
D}^2}$.  ``Linearity'' of $f$ would then give
\be
f(E)=\sum_{i=1}^{{\scriptscriptstyle D}^2} \alpha_i f(E_i)\;.
\ee
So, if we define $\rho$ by solving the $D^2$ linear equations
\be
\tr(\rho E_i)=f(E_i)\;,
\ee
we would have
\be
f(E)=\sum_i \alpha_i \tr\big(\rho E_i\big) = \tr\!\left(\rho\sum_i
\alpha_i E_i\right)= \tr(\rho E)
\label{HollyHobie}
\ee
and essentially be done.  (Positivity and normalization of $f$ would
require $\rho$ to be an actual density operator.)  But the {\it
problem\/} is that in expansion~(\ref{Hortense}) there is no
guarantee that the coefficients $\alpha_i$ can be chosen so that
$\alpha_i E_i\in{\cal E}_{\scriptscriptstyle\rm D}$.

What remains to be shown is that $f$ can be extended uniquely to a
function that is truly linear on ${\cal O}_{\scriptscriptstyle\rm
D}$.  This too is rather simple.  First, take any positive
semi-definite operator $E$. We can always find a positive rational
number $g$ such that $E=gG$ and $G\in{\cal E}_{\scriptscriptstyle\rm
D}$.  Therefore, we can simply define $f(E)\equiv g f(G)$.  To see
that this definition is unique, suppose there are two such operators
$G_1$ and $G_2$ (with corresponding numbers $g_1$ and $g_2$) such
that $E=g_1G_1=g_2G_2$. Further suppose $g_2\ge g_1$. Then
$G_2=\frac{g_1}{g_2}G_1$ and, by the homogeneity of the original
unextended definition of $f$, we obtain $g_2 f(G_2) = g_1 f(G_1)$.
Furthermore this extension retains the additivity of the original
function.  For suppose that neither $E$ nor $G$, though positive
semi-definite, are necessarily in ${\cal E}_{\scriptscriptstyle\rm
D}$.  We can find a positive rational number $c\ge1$ such that
$\frac{1}{c}(E+G)$, $\frac{1}{c}E$, and $\frac{1}{c}G$ are all in
${\cal E}_{\scriptscriptstyle\rm D}$.  Then, by the rules we have
already obtained,
\be
f(E+G)=c\, f\!\left(\frac{1}{c}(E+G)\right)= c \,
f\!\left(\frac{1}{c}E\right)+c\,f\!\left(\frac{1}{c}G\right)=
f(E)+f(G).
\ee

Let us now further extend $f$'s domain to the full space ${\cal
O}_d$. This can be done by noting that any operator $H$ can be
written as the difference $H=E-G$ of two positive semi-definite
operators.  Therefore define $f(H)\equiv f(E)-f(G)$, from which it
also follows that $f(-G)=-f(G)$. To see that this definition is
unique suppose there are four operators $E_1$, $E_2$, $G_1$, and
$G_2$, such that $H=E_1-G_1=E_2-G_2$.  It follows that
$E_1+G_2=E_2+G_1$. Applying $f$ (as extended in the previous
paragraph) to this equation, we obtain $f(E_1)+f(G_2)=f(E_2)+f(G_1)$
so that $f(E_1)-f(G_1)=f(E_2)-f(G_2)$. Finally, with this new
extension, full linearity can be checked immediately. This completes
the proof as far as the (complex) rational number field is
concerned:  Because $f$ extends uniquely to a linear functional on
${\cal O}_{\scriptscriptstyle\rm D}$, we can indeed go through the
steps of Eqs.~(\ref{Hortense}) through (\ref{HollyHobie}) without
worry. \normalsize
\eq

There are two things that are significant about this much of the
proof.  First, in contrast to Gleason's original theorem, there is
nothing to bar the same logic from working when $D=2$.  This is quite
nice because much of the community has gotten into the habit of
thinking that there is nothing particularly ``quantum mechanical''
about a single qubit.\,\cite{vanEnk00}  Indeed, because orthogonal
projectors on ${\cal H}_2$ can be mapped onto antipodes of the Bloch
sphere, it is known that the measurement-outcome statistics for any
standard measurement can be mocked-up through a noncontextual
hidden-variable theory.  What this result shows is that that simply
is not the case when one considers the full set of POVMs as one's
potential measurements.\,\footnote{In fact, one need not consider the
full set of POVMs in order to derive a noncolorability result along
the lines of Kochen and Specker for a single qubit.  Considering only
3-outcome POVMs of the so-called ``trine'' or ``Mercedes-Benz'' type
already does the trick.\,\cite{FuchsRenesOneDay}}

The other important thing is that the theorem works for Hilbert
spaces over the rational number field:  one does not need to invoke
the full power of the continuum. This contrasts with the surprising
result of Meyer\,\cite{Meyer99} that the standard Gleason theorem
fails in such a setting.  The present theorem hints at a kind of
resiliency to the structure of quantum mechanics that falls through
the mesh of the standard Gleason result: The probability rule for
POVMs does not actually depend so much upon the detailed workings of
the number field.

The final step of the proof, indeed, is to show that nothing goes
awry when we go the extra step of reinstating the continuum.
\begin{quote}
In other words, we need to show that the function $f$ (now defined
on the set ${\cal E}_{\scriptscriptstyle\rm D}$ complex operators)
is a continuous function. This comes about in simple way from $f$'s
additivity. Suppose for two positive semi-definite operators $E$ and
$G$ that $E\le G$ (i.e., $G-E$ is positive semi-definite). Then
trivially there exists a positive semi-definite operator $H$ such
that $E+H=G$ and through which the additivity of $f$ gives $f(E)\le
f(G)$.  Let $c$ be an irrational number, and let $a_n$ be an
increasing sequence and $b_n$ a decreasing sequence of rational
numbers that both converge to $c$. It follows for any positive
semi-definite operator $E$, that
\be
f(a_n E)\le f(cE) \le f(b_n E)\;,
\ee
which implies
\be
a_n f(E)\le f(cE) \le b_n f(E)\;.
\ee

Since $\lim a_n f(E)$ and $\lim b_n f(E)$ are identical, by the
``pinching theorem'' of elementary calculus, they must equal
$f(cE)$. This establishes that we can consistently define
\be
f(cE)=cf(E)\;.
\ee
Reworking the extensions of $f$ in the last inset (but with this
enlarged notion of homogeneity), one completes the proof in a
straightforward manner.
\end{quote}

Of course we are not getting something from nothing.  The reason the
present derivation is so easy in contrast to the standard proof is
that {\it mathematically\/} the assumption of POVMs as the basic
notion of measurement is significantly stronger than the usual
assumption.  {\it Physically}, though, I would say it is just the
opposite.  Why add extra restrictions to the notion of measurement
when they only make the route from basic assumption to practical
usage more circuitous than need be?

Still, no assumption should be left unanalyzed if it stands a chance
of bearing fruit.  Indeed, one can ask what is so very compelling
about the noncontextuality property (of probability assignments) that
both Gleason's original theorem and the present version make use of.
Given the picture of measurement as a kind of invasive intervention
into the world, one might expect the very opposite. One is left
wondering why measurement probabilities do not depend upon the whole
context of the measurement interaction.  Why is $P(d)$ not of the
form $f(d, \{E_c\})$?  Is there any good reason for this kind of
assumption?

\subsection{Noncontextuality}

In point of fact, there is: For, one can argue that the
noncontextuality of probability assignments for measurement outcomes
is more basic than even the particular structure of measurements
(i.e., that they be POVMs).  Noncontextuality bears more on how we
identify what we are measuring than anything to do with a
measurement's invasiveness upon nature.

Here is a way to see that.\,\cite{Mackey63}  Forget about quantum
mechanics for the moment and consider a more general world---one
that, skipping the details of quantum mechanics, still retains the
notions of systems, machines, actions, and consequences, and, most
essentially, retains the notion of a scientific agent performing
those actions and taking note of those consequences.

Take a system $S$ and imagine acting on it with one of two machines,
$M$ and $N$---things that we might colloquially call ``measurement
devices'' if we had the aid of a theory like quantum mechanics.  For
the case of machine $M$, let us label the possible consequences of
that action $\{ m_1, m_2, \ldots \}$. (Or if you want to think of
them in the mold of quantum mechanics, call them ``measurement
outcomes.'') For the case of machine $N$, let us label them $\{ n_1,
n_2, \ldots \}$.

If one takes a Bayesian point of view about probability, then nothing
can stop the agents in this world from using all the information
available to them to ascribe probabilities to the consequences of
those two potential actions. Thus, for an agent who cares to take
note, there are two probability distributions, $p_M(m_k)$ and
$p_N(n_k)$, lying around.  These probability distributions stand for
{\it his\/} subjective judgments about what will obtain if he acts
with either of the two machines.

This is well and good, but it is hardly a physical theory.  We need
more.  Let us suppose the labels $m_k$ and $n_k$ are, at the very
least, to be identified with elements in some master set $\cal
F$---that is, that there is some kind of connective glue for
comparing the operation of one machine to another. This set may even
be a set with further structure, like a vector space or something,
but that is beside the point. What is of first concern is under what
conditions will an agent identify two particular labels $m_i$ and
$n_j$ with the same element $F$ in the master set---disparate in
appearance, construction, and history though the two machines $M$
and $N$ may be. Perhaps one machine was manufactured by Lucent
Technologies while the other was manufactured by IBM Corporation.

There is really only one tool available for the purpose, namely the
probability assignments $p_M(m_i)$ and $p_N(n_j)$.  If
\be
p_M(m_i) \ne p_N(n_j)\;,
\ee
then surely he would not imagine identifying $m_i$ and $n_j$ with the
same element $E\in{\cal E}$. If, on the other hand, he finds
\be
p_M(m_i)=p_N(n_j)
\ee
{\it regardless\/} of his initial beliefs about about $S$, then we
might think there is some warrant for it.

That is the whole story of noncontextuality.  It is nothing more
than:  The consequences ($m_i$ and $n_j$) of our disparate actions
($M$ and $N$) should be labeled the same when we would bet the same
on them in all possible circumstances (i.e., regardless of our
initial knowledge of $S$).  To put this maybe a bit more baldly, the
label by which we identify a measurement outcome is a subjective
judgment just like a probability, and just like a quantum state.

By this point of view, noncontextuality is a tautology---it is built
in from the start.  Asking why we have it is a waste of time.  Where
we do have a freedom is in asking why we make one particular choice
of a master set over another.  Asking that may tell us something
about physics.  Why should the $m_i$'s be drawn from a set of effects
${\cal E}_{\scriptscriptstyle\rm D}$? Not all choices of the master
set are equally interesting once we have settled on noncontextuality
for the probability assignments.\footnote{See Ref.~\cite{Fuchs01a},
pp.~86--88, and Ref.~\cite{Baytch02} for some examples in that
regard.}  But quantum mechanics, of course, is particularly
interesting!

\subsection{Le Bureau International des Poids et Mesures \`a Paris}

There is still one further, particularly important, advantage to
thinking of POVMs as the basic notion of measurement in quantum
mechanics. For with an appropriately chosen {\it single\/} POVM one
can stop thinking of the quantum state as a linear operator
altogether, and instead start thinking of it as a probabilistic
judgment with respect to the (potential) outcomes of a {\it standard
quantum measurement}.  That is, a measurement device right next to
the standard kilogram and the standard meter in a carefully guarded
vault, deep within the bowels of the International Bureau of Weights
and Measures.\footnote{This idea has its roots in L.~Hardy's two
important papers Refs.~\cite{Hardy01a} and \cite{Hardy01b}.} Here is
what I mean by this.

Our problem hinges on finding a measurement for which the
probabilities of outcomes completely specify a unique density
operator.  Such measurements are called {\it informationally
complete\/} and have been studied for some
time~\cite{Prugovecki77,Schroeck91,Busch91}.  Here however, the
picture is most pleasing if we consider a slightly refined version
of the notion---that of the {\it minimal\/} informationally complete
measurement~\cite{Caves01}.  The space of Hermitian operators on
${\cal H}_{\rm\scriptscriptstyle D}$ is itself a linear vector space
of dimension $D^2$. The quantity ${\rm tr}(A^\dagger B)$ serves as
an inner product on that space. Hence, if we can find a POVM ${\cal
E}= \{E_d\}$ consisting of $D^2$ linearly independent operators, the
probabilities $P(d)={\rm tr}(\rho E_d)$---now thought of as {\it
projections\/} in the directions of the vectors $E_d$---will
completely specify the operator $\rho$. Any two distinct density
operators $\rho$ and $\sigma$ must give rise to distinct outcome
statistics for this measurement.  The minimal number of outcomes a
POVM can have and still be informationally complete is $D^2$.

Do minimal informationally complete POVMs exist?  The answer is yes.
Here is a simple way to produce one, though there are many other
ways. Start with a complete orthonormal basis $|e_j\rangle$ on ${\cal
H}_{\rm\scriptscriptstyle D}$.  It is easy to check that the
following $D^2$ rank-1 projectors $\Pi_d$ form a linearly
independent set.
\begin{enumerate}
\item For $d=1,\ldots,D$, let
\begin{equation}
\Pi_d = |e_j\rangle\langle e_j|\,,
\end{equation}
where $j$, too, runs over the values $1,\ldots,D$.

\item For $d=D+1,\ldots,\frac{1}{2}D(D+1)$, let
\begin{equation}
\Pi_d = \frac{1}{2}\big(|e_j\rangle+|e_k\rangle\big) \big(\langle
e_j|+\langle e_k|\big)\;,
\end{equation}
where $j<k$.

\item Finally, for $d= \frac{1}{2}D(D+1) + 1, \ldots,D^2$, let
\begin{equation}
\Pi_d = \frac{1}{2}\big(|e_j\rangle+i|e_k\rangle\big) \big(\langle
e_j|-i\langle e_k |\big)\;,
\end{equation}
where again $j<k$.
\end{enumerate}
All that remains is to transform these (positive-semidefinite)
linearly independent operators $\Pi_d$ into a proper POVM\@. This can
be done by considering the positive semidefinite operator $G$
defined by
\begin{equation}
G=\sum_{d=1}^{D^2}\Pi_d\;.
\label{Herbert}
\end{equation}
It is straightforward to show that $\langle\psi|G|\psi\rangle>0$ for
all $|\psi\rangle\ne0$, thus establishing that $G$ is positive
definite (i.e., Hermitian with positive eigenvalues) and hence
invertible.  Applying the (invertible) linear transformation
$X\rightarrow\, G^{-1/2}XG^{-1/2}$ to Eq.~(\ref{Herbert}), we find a
valid decomposition of the identity,
\begin{equation}
I=\sum_{d=1}^{D^2}G^{-1/2}\Pi_d G^{-1/2}\;.
\end{equation}
The operators
\begin{equation}
E_d=G^{-1/2}\Pi_d G^{-1/2}
\label{SquelchBuster}
\end{equation}
satisfy the conditions of a POVM, Eqs.~(\ref{Hank}) and
(\ref{Hannibal}), and moreover, they retain the rank and linear
independence of the original $\Pi_d$.  Thus we have what we need.

With the existence of minimal informationally complete POVMs
assured, we can think about the vault in Paris.  Let us suppose from
here out that it contains a machine that enacts a minimal
informationally complete POVM $E_h$ whenever it is used.  We reserve
the index $h$ to denote the outcomes of this standard quantum
measurement, for they will replace the notion of the ``hypothesis''
in classical statistical theory. Let us develop this from a Bayesian
point of view.

Whenever one has a quantum system in mind, it is legitimate for him
to use all he knows and believes of it to ascribe a probability
function $P(h)$ to the (potential) outcomes of this standard
measurement. In fact, that is all a quantum state {\it is\/} from
this point of view:  It is a subjective judgment about which
consequence will obtain as a result of an interaction between {\it
one's\/} system and {\it that\/} machine.  Whenever one performs a
measurement $\{E_d\}$ on the system---one different from the standard
quantum measurement $\{E_h\}$---at the most basic level of
understanding, all one is doing is gathering (or evoking) a piece of
data $d$ that (among other things) allows one to update from one's
initial subjective judgment $P(h)$ to {\it something\/} else
$P_d(h)$.\footnote{We will come back to describing the precise form
of this update and its similarity to Bayes' rule in Section 6.}

What is important to recognize is that, with this change of
description, we may already be edging toward a piece of quantum
mechanics that is not of information theoretic origin. It is this. If
one accepts quantum mechanics and supposes that one has a system for
which the standard quantum measurement device has $D^2$ outcomes
(for some integer $D$), then one is no longer completely free to
make just any subjective judgment $P(h)$ he pleases. There are
constraints.  Let us call the allowed region of initial judgments
${\cal P}_{\rm\scriptscriptstyle SQM}$.

\begin{figure} 
\begin{center}
\includegraphics[height=2in]{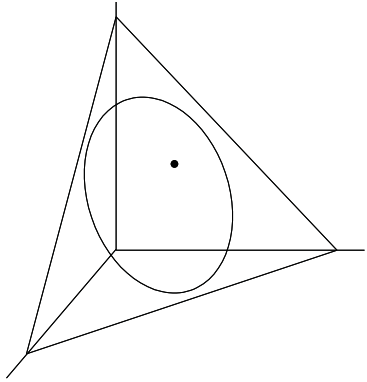}
\bigskip\caption{The planar surface represents the space of all
probability distributions over $D^2$ outcomes.  Accepting quantum
mechanics is, in part, accepting that one's subjective beliefs for
the outcomes of a standard quantum measurement device will not fall
outside a certain convex set. Each point within the region
represents a perfectly valid quantum state.}
\end{center}
\end{figure}

For instance, take the POVM in Eq.~(\ref{SquelchBuster}) as the
standard quantum measurement.  (And thus, now label its outcomes by
$h$ rather than $d$.)  Then, one can show that $P(h)$ is bounded
away from unity, regardless of one's initial quantum state for the
system.  In particular,
\bea
P(h) &\le& \max_\rho\,\tr(\rho E_h)
\nonumber
\\
&\le& \max_\Pi\,\tr(\Pi E_h)
\nonumber
\\
&\le& \lambda_{\rm max}(E_h)
\nonumber
\\
&=& \lambda_{\rm max}\big(G^{-1/2}\Pi_h G^{-1/2}\big)=
\lambda_{\rm max}\big(\Pi_h G^{-1}\Pi_h\big)
\nonumber
\\
&\le& \lambda_{\rm max}\big(G^{-1}\big)\;,
\eea
where the second line above refers to a maximization over all one
dimensional projectors and $\lambda_{\rm max}({\bf \cdot})$ denotes
the largest eigenvalue of its argument.  On the other hand, one can
calculate the eigenvalues of $G^{-1}$
explicitly.\,\cite{FuchsHuelsbergenPlunk} Through this, one obtains
\be
P(h)\le\left[D-\frac{1}{2}\!\left(1+\cot\frac{3\pi}{4D}\right)\right]^{-1}<1\;.
\ee
For large $D$, this bound asymptotes to roughly $(0.79 D)^{-1}$.

More generically, for any minimal informationally complete POVM
$\{E_h\}$, $P(h)$ must be bounded away from unity for all its
possible outcomes. Thus even at this stage, there is something
driving a wedge between quantum mechanics and simple Bayesian
probability theory.  When one accepts quantum mechanics, one
voluntarily accepts a restriction on one's subjective judgments for
the consequences of a standard quantum measurement intervention: For
all consequences $h$, there are no conditions whatsoever convincing
enough to compel one to a probability ascription $P(h)=1$.  That is,
one gives up on the hope of certainty. This, indeed, one might
pinpoint as an assumption about the physical world that goes beyond
pure probability theory.\footnote{It is at this point that the
present account of quantum mechanics differs most crucially from
Refs.~\cite{Hardy01a} and \cite{Hardy01b}\@.  Hardy sees quantum
mechanics as a generalization and extension of classical probability
theory, whereas quantum mechanics is depicted here as a restriction
to probability theory. It is a restriction that takes into account
how we ought to think and gamble in light of a certain physical
fact---a fact we are working like crazy to identify.}

But what is that assumption in physical terms?  What is our best
description of the wedge? Some think they already know the answer,
and it is quantum entanglement.

\section{Wither Entanglement?\,\protect\footnote{This is not a
spelling mistake.}}
\begin{flushright}
\parbox{4.18in}{\footnotesize
\bq
When two systems, of which we know the states by their respective
representatives, enter into temporary physical interaction due to
known forces between them, and when after a time of mutual influence
the systems separate again, then they can no longer be described in
the same way as before, viz.\ by endowing each of them with a
representative of its own.  I would not call that {\it one\/} but
rather {\it the\/} characteristic trait of quantum mechanics, the one
that enforces its entire departure from classical lines of thought.
By the interaction the two representatives (or $\psi$-functions) have
become entangled.\\
\hspace*{\fill} --- {\it Erwin Schr\"odinger}, 1935
\eq
}
\end{flushright}

Quantum entanglement has certainly captured the attention of our
community. By most accounts it is the main ingredient in quantum
information theory and quantum computing\,\cite{Jozsa98}, and it is
the main mystery of the quantum foundations\,\cite{Mermin90}. But
what is it? Where does it come from?

The predominant purpose it has served in this paper has been as a
kind of background.  For it, more than any other ingredient in
quantum mechanics, has clinched the issue of ``information about
what?''~in the author's mind:  That information cannot be about a
preexisting reality (a hidden variable) unless we are willing to
renege on our reason for rejecting the quantum state's objective
reality in the first place. What I am alluding to here is the
conjunction of the Einstein argument reported in Section 3 and the
phenomena of the Bell inequality violations by quantum mechanics.
Putting those points together gave us that the information
symbolized by a $|\psi\rangle$ must be information about the
potential consequences of our interventions into the world.

But, now I would like to turn the tables and ask whether the
structure of our potential interventions---the POVMs---can tell us
something about the origin of entanglement.  Could it be that the
concept of entanglement is just a minor addition to the much deeper
point that measurements have this structure?

The technical translation of this question is, why do we combine
systems according to the tensor-product rule?  There are certainly
innumerable ways to combine two Hilbert spaces ${\cal H}_A$ and
${\cal H}_B$ to obtain a third ${\cal H}_{AB}$.  We could take the
direct sum of the two spaces ${\cal H}_{AB}={\cal H}_{A}\oplus{\cal
H}_{B}$. We could take their Grassmann product ${\cal H}_{AB}={\cal
H}_{A}\wedge{\cal H}_{B}$\,\cite{Bhatia97}.  We could take scads of
other things. But instead we take their tensor product,
\be
{\cal H}_{AB}={\cal H}_{A}\otimes{\cal H}_{B}\;.
\ee
Why?

Could it arise from the selfsame considerations as of the previous
section---namely, from a noncontextuality property for
measurement-outcome probabilities? The answer is yes, and the
theorem I am about demonstrate owes much in inspiration to
Ref.~\cite{Wallach00}.\,\footnote{After posting Ref.~\cite{Fuchs01b},
Howard Barnum and Alex Wilce brought to my attention that there is a
significant amount of literature in the quantum logic community
devoted to similar ways of motivating the tensor-product rule. See
for example Ref.~\cite{Aerts78} and the many citations therein.}

Here is the scenario. Suppose we have two quantum systems, and we can
make a measurement on each.  On the first, we can measure any POVM on
the $D_{\rm A}$-dimensional Hilbert space ${\cal H}_A$; on the
second, we can measure any POVM on the $D_{\rm B}$-dimensional
Hilbert space ${\cal H}_B$. (This, one might think, is the very
essence of having {\it two\/} systems rather than one---i.e., that we
can probe them independently.)  Moreover, suppose we may condition
the second measurement on the nature and the outcome of the first,
and vice versa. That is to say---walking from $A$ to $B$---we could
first measure $\{E_i\}$ on $A$, and then, depending on the outcome
$i$, measure $\{F^i_j\}$ on $B$. Similarly---walking from $B$ to
$A$---we could first measure $\{F_j\}$ on $B$, and then, depending
on the outcome $j$, measure $\{E^j_i\}$ on $A$. So that we have valid
POVMs, we must have
\be
\sum_i E_i = I \qquad\mbox{and}\qquad \sum_j F^i_j = I \quad\forall\,
i\;,
\label{Herme}
\ee
and
\be
\sum_i E^j_i = I \quad\forall\, j \qquad\mbox{and}\qquad \sum_j F_j =
I\;,
\label{Neutics}
\ee
for these sets of operators.  Let us denote by $S_{ij}$ an {\it
ordered pair\/} of operators, either of the form $(E_i, F^i_j)$ or
of the form $(E^j_i, F_j)$, as appearing above.  Let us call a set
of such operators $\{S_{ij}\}$ a {\it locally-measurable POVM tree}.

Suppose now that---just as with the POVM-version of Gleason's theorem
in Section 4---the joint probability $P(i,j)$ for the outcomes of
such a measurement should not depend upon which tree $S_{ij}$ is
embedded in:  This is essentially the same assumption we made there,
but now applied to local measurements on the separate systems. In
other words, let us suppose there exists a function
\be
f:{\cal E}_{\scriptscriptstyle\rm D_A}\!\times {\cal
E}_{\scriptscriptstyle\rm D_B} \longrightarrow [0,1]
\label{Ersatz}
\ee
such that
\be
\sum_{ij} f(S_{ij}) = 1
\label{HamAndEggs}
\ee
whenever the $S_{ij}\/$ satisfy either Eq.~(\ref{Herme}) or
Eq.~(\ref{Neutics}).

Note in particular that Eq.~(\ref{Ersatz}) makes no use of the
tensor product:  The domain of $f$ is the {\it Cartesian product\/}
of the two sets ${\cal E}_{\scriptscriptstyle\rm D_A}$ and ${\cal
E}_{\scriptscriptstyle\rm D_B}$.  The notion of a {\it local\/}
measurement on the separate systems is enforced by the requirement
that the ordered pairs $S_{ij}$ satisfy the side conditions of
Eqs.~(\ref{Herme}) and (\ref{Neutics}).  This, of course, is not the
most general kind of local measurement one can imagine---more
sophisticated measurements could involve multiple ping-pongings
between $A$ and $B$ as in Ref.~\cite{Bennett99}---but the present
restricted class is already sufficient for fixing that the
probability rule for local measurements must come from a
tensor-product structure.

The {\it theorem}\,\footnote{In Ref.~\cite{Fuchs01b}, a
significantly stronger claim is made:  Namely, that $\cal L$ is in
fact a density operator.  This was a flat-out mistake. See further
discussion below.} is this: If $f$ satisfies Eqs.~(\ref{Ersatz}) and
(\ref{HamAndEggs}) for all locally-measurable POVM trees, then there
exists a linear operator $\cal L$ on ${\cal H}_{A}\otimes{\cal
H}_{B}$ such that
\be
f(E,F)=\tr\Big({\cal L}(E\otimes F)\Big)\;.
\ee
If ${\cal H}_{A}$ and ${\cal H}_{B}$ are defined over the field of
complex numbers, then $\cal L$ is unique.  Uniqueness does not hold,
however, if the underlying field is the real numbers.

The proof of this statement is almost a trivial extension of the
proof in Section 4.  One again starts by showing additivity, but
this time in the two variables $E$ and $F$ separately.  For
instance, for a fixed $E\in{\cal E}_{\scriptscriptstyle\rm D_A}$,
define
\be
g_E(F) = f(E,F)\;,
\ee
and consider two locally-measurable POVM trees
\be
\{(I-E,F_i), (E, G_\alpha)\} \qquad\mbox{and}\qquad \{(I-E,F_i), (E,
H_\beta)\}\;,
\ee
where $\{F_i\}$, $\{G_\alpha\}$, and $\{H_\beta\}$ are arbitrary
POVMs on ${\cal H}_B$.  Then Eq.~(\ref{HamAndEggs}) requires that
\be
\sum_i g_{I\mbox{-}E}(F_i)+\sum_\alpha g_E(G_\alpha) = 1
\ee
and
\be
\sum_i g_{I\mbox{-}E}(F_i)+\sum_\beta g_E(H_\beta) = 1\;.
\ee
From this it follows that,
\be
\sum_\alpha g_E(G_\alpha) = \sum_\beta g_E(H_\beta) = \mbox{const}.
\ee
That is to say, $g_E(F)$ is a frame function in the sense of Section
4.  Consequently, we know that we can use the same methods as there
to uniquely extend $g_E(F)$ to a linear functional on the complete
set of Hermitian operators on ${\cal H}_B$. Similarly, for fixed
$F\in{\cal E}_{\scriptscriptstyle\rm D_B}$, we can define
\be
h_F(E) = f(E,F)\;,
\ee
and prove that this function too can be extended uniquely to a linear
functional on the Hermitian operators on ${\cal H}_A$.

The linear extensions of $g_E(F)$ and $h_F(E)$ can be put together
in a simple way to give a full bilinear extension to the function
$f(E,F)$.  Namely, for any two Hermitian operators $A$ and $B$ on
${\cal H}_A$ and ${\cal H}_B$, respectively, let $A=\alpha_1 E_1 -
\alpha_2 E_2$ and $B=\beta_1 F_1 - \beta_2 F_2$ be decompositions
such that $\alpha_1,\alpha_2,\beta_1,\beta_2\ge0$, $E_1,E_2\in {\cal
E}_{D_{\scriptscriptstyle\rm A}}$, and $F_1,F_2\in {\cal
E}_{D_{\scriptscriptstyle\rm B}}$.  Then define
\be
f(A,B)
\nonumber
\\
\equiv \alpha_1 g_{E_1}(B) - \alpha_2 g_{E_2}(B)\;.
\ee
To see that this definition is unique, take any other decomposition
\be
A=\tilde\alpha_1 \tilde E_1 - \tilde\alpha_2 \tilde E_2\;.
\ee
Then we have
\bea
f(A
,B)
&=&
\tilde \alpha_1 g_{\tilde E_1}(B) -
\tilde \alpha_2 g_{\tilde E_2}(B)
\nonumber
\\
&=&
\tilde \alpha_1 f(\tilde E_1,B) - \tilde \alpha_2 f(\tilde E_2,B)
\nonumber
\\
&=&
\beta_1 \Big(\tilde \alpha_1 f(\tilde E_1,F_1)- \tilde
\alpha_2 f(\tilde E_1,F_1)\Big) -
\beta_2 \Big(\tilde \alpha_1 f(\tilde E_1,F_2)- \tilde
\alpha_2 f(\tilde E_2,F_2)\Big)
\nonumber
\\
&=&
\beta_1 h_{F_1}(A) - \beta_2 h_{F_2}(A)
\nonumber
\\
&=&
\beta_1 \Big(\alpha_1 f(E_1,F_1)- \alpha_2 f(E_1,F_1)\Big) -
\beta_2 \Big(\alpha_1 f(E_1,F_2)- \alpha_2 f(E_2,F_2)\Big)
\nonumber
\\
&=&
\alpha_1 f(E_1,B) - \alpha_2 f(E_2,B)
\nonumber
\\
&=&
\alpha_1 g_{E_1}(B) - \alpha_2 g_{E_2}(B)\;,
\eea
which is as desired.

With bilinearity for the function $f$ established, we have
essentially the full story\,\cite{Bhatia97,MacLane67}.  For, let
$\{E_i\}$, $i=1,\ldots,D_{\rm A}^2$, be a complete basis for the
Hermitian operators on ${\cal H}_A$ and let $\{F_j\}$,
$j=1,\ldots,D_{\rm B}^2$, be a complete basis for the Hermitian
operators on ${\cal H}_B$.  If $E=\sum_i\alpha_i E_i$ and
$F=\sum_j\beta_j F_j$, then
\be
f(E,F)=\sum_{ij}\alpha_i\beta_j f(E_i,F_j)\;.
\ee
Define $\cal L$ to be a linear operator on ${\cal H}_{A}\otimes{\cal
H}_{B}$ satisfying the $(D_{\rm A}D_{\rm B})^2$ linear equations
\be
\tr\Big({\cal L}(E_i\otimes F_j)\Big)=f(E_i,F_j)\;.
\label{BigGrump}
\ee
Such an operator always exists.  Consequently we have,
\bea
f(E,F) &=& \sum_{ij}\alpha_i\beta_j \tr\Big({\cal L}(E_i\otimes
F_j)\Big)
\nonumber
\\
&=&
\tr\Big({\cal L}(E\otimes F)\Big)\;.
\eea

For complex Hilbert spaces ${\cal H}_A$ and ${\cal H}_B$, the
uniqueness of ${\cal L}$ follows because the set $\{E_i\otimes F_j\}$
forms a complete basis for the Hermitian operators on ${\cal
H}_{A}\otimes{\cal H}_{B}$.\,\cite{Araki80}  For real Hilbert spaces,
however, the analog of the Hermitian operators are the symmetric
operators. The dimensionality of the space of symmetric operators on
a real Hilbert space ${\cal H}_{\scriptscriptstyle\rm D}$ is
$\frac{1}{2}D(D+1)$, rather than the $D^2$ it is for the complex
case.  This means that in the steps above only
\be
\frac{1}{4} D_{\rm A}D_{\rm B}(D_{\rm A}+1)(D_{\rm B}+1)
\label{LoudTalk}
\ee
equations will appear in Eq.~(\ref{BigGrump}), whereas
\be
\frac{1}{2} D_{\rm A}D_{\rm B} (D_{\rm A}D_{\rm B}+1)
\label{Ulcer}
\ee
are needed to uniquely specify an ${\cal L}$.  For instance take
$D_{\rm A}=D_{\rm B}=2$.  Then Eq.~(\ref{LoudTalk}) gives nine
equations, while Eq.~(\ref{Ulcer}) requires ten.

This establishes the theorem.  It would be nice if we could go
further and establish the full probability rule for local quantum
measurements---i.e., that ${\cal L}$ must be a density operator.
Unfortunately, our assumptions are not strong enough for that.  Here
is a counterexample.\,\cite{Wilce90}  Consider a linear operator that
is proportional to the swap operator on the two Hilbert spaces:
\be
{\cal L}_{\rm\scriptscriptstyle S}(E\otimes
F)=\frac{1}{D^{\,2}}\,F\otimes E\;.
\ee
This clearly satisfies the conditions of our theorem, but it is not
equivalent to a density operator.

Of course, one could recover positivity for ${\cal L}$ by requiring
that it give positive probabilities even for nonlocal measurements
(i.e., resolutions of the identity operator on ${\cal H}_{A}\otimes
{\cal H}_{B}$)\@.  But in the purely local setting contemplated here,
that would be a cheap way out.  For, one should ask in good
conscience what ought to be the rule for defining the full class of
measurements (including nonlocal measurements): Why should it
correspond to an arbitrary resolution of the identity on the tensor
product? There is nothing that makes it obviously so, unless one has
already accepted standard quantum mechanics.

Alternatively, it must be possible to give a purely local condition
that will restrict ${\cal L}$ to be a density operator.  This is
because ${\cal L}$, as noted above, is uniquely determined by the
function $f(E,F)$; we never need to look further than the
probabilities of local measurements outcomes in specifying ${\cal
L}$. Ferreting out such a condition supplies an avenue for future
research.

All of this does not, however, take away from the fact that whatever
${\cal L}$ is, it must be a linear operator on the tensor product of
${\cal H}_{A}$ and ${\cal H}_{B}$.  Therefore, let us close by
emphasizing the striking feature of this way of deriving the
tensor-product rule for combining separate quantum systems:  It is
built on the very concept of local measurement. There is nothing
``spooky'' or ``nonlocal'' about it; there is nothing in it
resembling ``passion at a distance''\,\cite{ShimonyFest}.  Indeed,
one did not even have to consider probability assignments for the
outcomes of measurements of the ``nonlocality without entanglement''
variety\,\cite{Bennett99} to uniquely fix the probability rule. That
is---to give an example on ${\cal H}_3\otimes{\cal H}_3$---one need
not consider standard measurements like
$\{E_d=|\psi_d\rangle\langle\psi_d|\}$, $d=1,\ldots,9$, where
\bea
|\psi_1\rangle \!\! &=& \!\! |1\rangle |1\rangle \nonumber\\
|\psi_2\rangle = |0\rangle |0+1\rangle \quad\qquad && \,\quad\qquad
|\psi_6\rangle = |1+2\rangle |0\rangle \nonumber\\
|\psi_3\rangle = |0\rangle |0-1\rangle \quad\qquad && \,\quad \qquad
|\psi_7\rangle = |1-2\rangle |0\rangle\\
|\psi_4\rangle = |2\rangle |1+2\rangle \quad\qquad && \,\quad \qquad
|\psi_8\rangle = |0+1\rangle |2\rangle \nonumber\\
|\psi_5\rangle = |2\rangle |1-2\rangle \quad\qquad && \,\quad \qquad
|\psi_9\rangle = |0-1\rangle |2\rangle \nonumber
\eea
with $|0\rangle$, $|1\rangle$, and $|2\rangle$ forming an
orthonormal basis on ${\cal H}_3$, and
$|0+1\rangle=\frac{1}{\sqrt{2}}(|0\rangle+|1\rangle)$, etc.  This is
a measurement that takes neither the form of Eq.~(\ref{Herme}) nor
(\ref{Neutics}).  It stands out instead, in that even though all its
POVM elements are tensor-product operators---i.e., they have no
quantum entanglement---it still {\it cannot\/} be measured by local
means, even with the elaborate ping-ponging strategies mentioned
earlier.

Thus, the tensor-product rule, and with it quantum entanglement,
seems to be more a statement of locality than anything else.  It,
like the probability rule, is more a product of the structure of the
observables---that they are POVMs---combined with noncontextuality.
In searching for the secret ingredient to drive a wedge between
general Bayesian probability theory and quantum mechanics, it seems
that the direction {\it not\/} to look is toward quantum
entanglement. Perhaps the trick instead is to dig deeper into the
Bayesian toolbox.

\section{Whither Bayes' Rule?\,\protect\footnote{This is not a
spelling mistake.}}
\begin{flushright}
\parbox{2.80in}{\footnotesize
\bq
And so you see I have come to doubt\\ All that I once held as true\\
I stand alone without beliefs\\ The only truth I know is you.\smallskip\\
\hspace*{\fill} --- {\it Paul Simon}, timeless
\eq
}
\end{flushright}

Quantum states are states of information, knowledge, belief,
pragmatic gambling commitments, {\it not\/} states of nature. That
statement is the cornerstone of this paper. Thus, in searching to
make sense of the remainder of quantum mechanics, one strategy ought
to be to seek guidance\,\cite{CavesFuchsSchack01} from the most
developed avenue of ``rational-decision theory'' to date---Bayesian
probability theory\,\cite{Kyburg80,JaynesPosthumous,Bernardo94}.
Indeed, the very aim of Bayesian theory is to develop reliable
methods of reasoning and making decisions in the light of incomplete
information.  To what extent does that structure mesh with the
seemingly independent structure of quantum mechanics? To what extent
are there analogies; to what extent distinctions?

This section is about turning a distinction into an analogy.  The
core of the matter is the manner in which states of belief are
updated in the two theories.  At first sight, they appear to be quite
different in character.  To see this, let us first explore how
quantum mechanical states change when information is gathered.

In older accounts of quantum mechanics, one often encounters the
``collapse postulate'' as a basic statement of the theory.  One hears
things like, ``Axiom 5:  Upon the completion of an ideal measurement
of an Hermitian operator $H$, the system is left in an eigenstate of
$H$.'' In quantum information, however, it has become clear that it
is useful to broaden the notion of measurement, and with it, the
analysis of how a state can change in the process.  The foremost
reason for this is that the collapse postulate is simply not true in
general:  Depending upon the exact nature of the measurement
interaction, there may be any of a large set of possibilities for
the final state of a system.

The broadest consistent notion of state change arises in the theory
of ``effects and operations''\,\cite{Kraus83}. The statement is this.
Suppose one's initial state for a quantum system is a density
operator $\rho$, and a POVM $\{E_d\}$ is measured on that system.
Then, according to this formalism, the state after the measurement
can be {\it any\/} state $\rho_d$ of the form
\be
\rho_d = \frac{1}{\tr(\rho E_d)}\sum_i A_{di}\rho A_{di}^\dagger\;,
\label{StinkFart}
\ee
where
\be
\sum_i A_{di}^\dagger A_{di} = E_d\;.
\label{ColdWarsawDay}
\ee
Note the immense generality of this formula.  There is no constraint
on the number of indices $i$ in the $A_{di}$ and these operators
need not even be Hermitian.

The {\it usual\/} justification for this kind of generality---just as
in the case of the commonplace justification for the POVM
formalism---comes about by imagining that the measurement arises in
an indirect fashion rather than as a direct and immediate
observation. In other words, the primary system is pictured to
interact with an ancilla first, and only then subjected to a
``real'' measurement on the ancilla alone. The trick is that one
posits a kind of projection postulate on the primary system due to
this process. This assumption has a much safer feel than the raw
projection postulate since, after the interaction, no measurement on
the ancilla should cause a physical perturbation to the primary
system.

More formally, we can start out by following Eqs.~(\ref{MushMush})
and (\ref{Aluminium}), but in place of Eq.~(\ref{RibTie}) we must
make an assumption on how the system's state changes.  For this one
invokes a kind of
``projection-postulate-at-a-distance.''\footnote{David Mermin has
also recently emphasized this point in Ref.~\cite{Mermin01}.} Namely,
one takes
\be
\rho_d=\frac{1}{P(d)}\;{\rm tr}_{\scriptscriptstyle\rm
A}\!\left((I\otimes\Pi_d) U(\rho_{\scriptscriptstyle\rm
S}\otimes\rho_{\scriptscriptstyle\rm A})
U^\dagger(I\otimes\Pi_d)\right)\;.
\label{BurpAtNight}
\ee
The reason for invoking the partial trace is to make sure that any
hint of a state change for the ancilla remains unaddressed.

To see how expression (\ref{BurpAtNight}) makes connection to
Eq.~(\ref{StinkFart}), denote the eigenvalues and eigenvectors of
$\rho_{\scriptscriptstyle\rm A}$ by $\lambda_\alpha$ and
$|a_\alpha\rangle$ respectively. Then $\rho_{\scriptscriptstyle\rm
S}\otimes\rho_{\scriptscriptstyle\rm A}$ can be written as
\be
\rho_{\scriptscriptstyle\rm S}\otimes\rho_{\scriptscriptstyle\rm
A}=\sum_\alpha
\sqrt{\lambda_\alpha}\,|a_\alpha\rangle\rho_{\scriptscriptstyle\rm
S} \langle a_\alpha|\sqrt{\lambda_\alpha}\;,
\ee
and, expanding Eq.~(\ref{BurpAtNight}), we have
\bea
\rho_d
&=&
\frac{1}{P(d)}\sum_\beta\langle a_\beta| (I\otimes\Pi_d) U^\dagger
(\rho_{\scriptscriptstyle\rm S}\otimes\rho_{\scriptscriptstyle\rm
A}) U (I\otimes\Pi_d)|a_\beta\rangle
\nonumber
\\
&=&
\frac{1}{P(d)}\sum_{\alpha\beta}\left(\sqrt{\lambda_\alpha}\,\langle
a_\beta| (I\otimes\Pi_d) U^\dagger
|a_\alpha\rangle\right)\rho_{\scriptscriptstyle\rm S}\left(\langle
a_\alpha| U(I\otimes\Pi_d)|a_\beta\rangle\sqrt{\lambda_\alpha}\,
\right). \rule{0mm}{8mm}
\nonumber
\\
&&
\eea
A representation of the form in Eq.~(\ref{StinkFart}) can be made by
taking
\be
A_{b\alpha\beta}=\sqrt{\lambda_\alpha}\,\langle a_\alpha|
U(I\otimes\Pi_d)|a_\beta\rangle
\label{GoToBed}
\ee
and lumping the two indices $\alpha$ and $\beta$ into the single
index $i$. Indeed, one can easily check that
Eq.~(\ref{ColdWarsawDay}) holds.\footnote{As an aside, it should be
clear from the construction in Eq.~(\ref{GoToBed}) that there are
many equally good representations of $\rho_d$.  For a precise
statement of the latitude of this freedom, see
Ref.~\cite{Schumacher96}.} This completes what we had set out to
show. However, just as with the case of the POVM $\{E_d\}$, one can
always find a way to reverse engineer the derivation:  Given a set of
$A_{di}$, one can always find a $U$, a $\rho_{\scriptscriptstyle\rm
A}$, and set of $\Pi_d$ such that Eq.~(\ref{BurpAtNight}) becomes
true.

Of course the old collapse postulate is contained within the extended
formalism as a special case:  There, one just takes both sets
$\{E_d\}$ and $\{A_{di}=E_d\}$ to be sets of orthogonal projection
operators. Let us take a moment to think about this special case in
isolation. What is distinctive about it is that it captures in the
extreme a common folklore associated with the measurement process.
For it tends to convey the image that measurement is a kind of
gut-wrenching violence: In one moment the state is
$\rho=|\psi\rangle\langle\psi|$, while in the very next it is a
$\Pi_i=|i\rangle\langle i|$. Moreover, such a wild transition need
depend upon no details of $|\psi\rangle$ and $|i\rangle$; in
particular the two states may even be almost orthogonal to each
other. In density-operator language, there is no sense in which
$\Pi_i$ is contained in $\rho$: the two states are in distinct
places of the operator space.  That is,
\be
\rho \ne \sum_i P(i) \Pi_i\;.
\ee

Contrast this with the description of information gathering that
arises in Bayesian probability theory.  There, an initial state of
belief is captured by a probability distribution $P(h)$ for some
hypothesis $H$.  The way gathering a piece of data $d$ is taken into
account in assigning one's new state of belief is through Bayes'
conditionalization rule.  That is to say, one expands $P(h)$ in terms
of the relevant joint probability distribution and picks off the
appropriate term:
\bea
P(h) &=& \sum_d P(h,d)
\nonumber
\\
&=& \sum_d P(d)P(h|d)
\label{Orecchiette}
\\
&& \phantom{\sum_d P(d)P(} \downarrow \nonumber\\
P(h) &&\stackrel{d}{\longrightarrow} \phantom{\sum_d} P(h|d)\;,
\label{Lasagna}
\eea
where $P(h|d)$ satisfies the tautology
\be
P(h|d)=\frac{P(h,d)}{P(d)}\;.
\label{Macaroni}
\ee
How gentle this looks in comparison to quantum collapse!  When one
gathers new information, one simply refines one's old beliefs in the
most literal of senses.  It is not as if the new state is
incommensurable with the old.  {\it It was always there}; it was
just initially averaged in with various other potential beliefs.

Why does quantum collapse not look more like Bayes' rule?  Is
quantum collapse really a more violent kind of change, or might it be
an artifact of a problematic representation?  By this stage, it
should come as no surprise to the reader that dropping the ancilla
from our image of generalized measurements will be the first step to
progress.  Taking the transition from $\rho$ to $\rho_d$ in
Eqs.~(\ref{StinkFart}) and (\ref{ColdWarsawDay}) as the basic
statement of what quantum measurement {\it is\/} is a good starting
point.

To accentuate a similarity between Eq.~(\ref{StinkFart}) and Bayes'
rule, let us first contemplate cases of it where the index $i$ takes
on a single value.  Then, we can conveniently drop that index and
write
\be
\rho_d = \frac{1}{P(d)}A_d\rho A_d^\dagger\;,
\label{SlopFart}
\ee
where
\be
E_d = A_d^\dagger A_d\;.
\label{GrayJerseyDay}
\ee
In a loose way, one can say that measurements of this sort are the
most efficient they can be for a given POVM $\{E_d\}$:  For, a
measurement interaction with an explicit $i$-dependence may be viewed
as ``more truly'' a measurement of a finer-grained POVM that just
happens to throw away some of the information it gained.   Let us
make this point more precise.

Notice that Bayes' rule has the property that one's uncertainty
about a hypothesis can be expected to decrease upon the acquisition
of data.  This can be made rigorous, for instance, by gauging
uncertainty with the Shannon entropy function \,\cite{Cover91},
\be
S(H) = - \sum_h P(h)\log P(h)\;.
\ee
This number is bounded between 0 and the logarithm of the number of
hypotheses in $H$, and there are several reasons to think of it as a
good measure of uncertainty. Perhaps the most important of these is
that it quantifies the number of binary-valued questions one expects
to ask (per instance of $H$) if one's only means to ascertain the
outcome is from a colleague who knows the result\,\cite{Ash65}.
Under this quantification, the lower the Shannon entropy, the more
predictable a measurement's outcomes.

Because the function $f(x)=-x\log x$ is concave on the interval
$[0,1]$, it follows that,
\bea
S(H) &=& - \sum_h \left( \sum_d P(d)P(h|d)\right)\log \left( \sum_d
P(d)P(h|d)\right)
\nonumber
\\
&\ge&
-\sum_d P(d)\sum_h P(h|d)\log P(h|d)\;.
\nonumber
\\
&=&
\sum_d P(d) S(H|d)
\label{HungryBelly}
\eea

Indeed we hope to find a similar statement for how the result of
efficient quantum measurements decrease uncertainty or
impredictability. But, what can be meant by a decrease of uncertainty
through quantum measurement? I have argued strenuously that the
information gain in a measurement cannot be information about a
preexisting reality.  The way out of the impasse is simple: The
uncertainty that decreases in quantum measurement is the uncertainty
one expects for the results of other potential measurements.

There are at least two ways of quantifying this that are worthy of
note. The first has to do with the von Neumann entropy of a density
operator $\rho$:
\be
S(\rho) = - {\rm tr}\,\rho\log\rho
=-\sum_{k=1}^D\lambda_k\log\lambda_k\;,
\label{Lahti}
\ee
where the $\lambda_k$ signify the eigenvalues of $\rho$.  (We use
the convention that $\lambda\log\lambda=0$ whenever $\lambda=0$ so
that $S(\rho)$ is always well defined.)

The intuitive meaning of the von Neumann entropy can be found by
first thinking about the Shannon entropy. Consider any von Neumann
measurement $\cal P$ consisting of $d$ one-dimensional orthogonal
projectors $\Pi_i$. The Shannon entropy for the outcomes of this
measurement is given by
\be
H({\cal P})=-\sum_{i=1}^D \big({\rm tr}\rho\Pi_i\big)\log\big({\rm
tr}\rho\Pi_i\big)\;.
\ee
A natural question to ask is:  With respect to a given density
operator $\rho$, which measurement $\cal P$ will give the most
predictability over its outcomes? As it turns out, the answer is any
$\cal P$ that forms a set of eigenprojectors for $\rho$
\,\cite{Wehrl78}. When this obtains, the Shannon entropy of the
measurement outcomes reduces to simply the von Neumann entropy of
the density operator. The von Neumann entropy, then, signifies the
amount of impredictability one achieves by way of a standard
measurement in a best case scenario.  Indeed, true to one's
intuition, one has the most predictability by this account when
$\rho$ is a pure state---for then $S(\rho)=0$.  Alternatively, one
has the least knowledge when $\rho$ is proportional to the identity
operator---for then any measurement $\cal P$ will have outcomes that
are all equally likely.

The best case scenario for predictability, however, is a limited
case, and not very indicative of the density operator as a whole.
Since the density operator contains, in principle, all that can be
said about every possible measurement, it seems a shame to throw
away the vast part of that information in our considerations.

This leads to a second method for quantifying uncertainty in the
quantum setting. For this, we again rely on the Shannon information
as our basic notion of impredictability. The difference is we
evaluate it with respect to a ``typical'' measurement rather than
the best possible one. But typical with respect to what? The notion
of typical is only defined with respect to a given {\it measure\/} on
the set of measurements.

Regardless, there is a fairly canonical answer. There is a unique
measure $d\Omega_\Pi$ on the space of one-dimensional projectors
that is invariant with respect to all unitary operations.  That in
turn induces a canonical measure $d\Omega_{\cal P}$ on the space of
von Neumann measurements $\cal P$ \,\cite{Wootters90}. Using this
measure leads to the following quantity
\begin{eqnarray}
\overline{S}(\rho)&=&\int H(\Pi)\,d\Omega_{\cal P}
\nonumber
\\
&=& -D \int \big({\rm tr}\rho\Pi\big)\log\big({\rm
tr}\rho\Pi\big)\,d\Omega_\Pi\;,
\end{eqnarray}
which is intimately connected to the so-called quantum
``subentropy'' of Ref.~\cite{Jozsa94}.  This mean entropy can be
evaluated explicitly in terms of the eigenvalues of $\rho$ and takes
on the expression
\be
\overline{S}(\rho)=\frac{1}{\ln2}\left(\frac{1}{2}+
\frac{1}{3}+\cdots+\frac{1}{D}\right)+ Q(\rho)
\ee
where the subentropy $Q(\rho)$ is defined by
\be
Q(\rho)=-\sum_{k=1}^D\! \left(\prod_{i\ne
k}\frac{\lambda_k}{\lambda_k-\lambda_i}\right)\!\lambda_k\log\lambda_k\;.
\label{Mittelstaedt}
\ee
In the case where $\rho$ has degenerate eigenvalues,
$\lambda_l=\lambda_m$ for $l\ne m$, one need only reset them to
$\lambda_l+\epsilon$ and $\lambda_m-\epsilon$ and consider the limit
as $\epsilon\rightarrow0$.  The limit is convergent and hence
$Q(\rho)$ is finite for all $\rho$.  With this, one can also see
that for a pure state $\rho$, $Q(\rho)$ vanishes. Furthermore, since
$\overline{S}(\rho)$ is bounded above by $\log d$, we know that
\be
0\le Q(\rho)\le\log d - \frac{1}{\ln2}\!\left(\frac{1}{2}+
\cdots+\frac{1}{D}\right)\le\frac{1-\gamma}{\ln 2}\;,
\ee
where $\gamma$ is Euler's constant.  This means that for any $\rho$,
$Q(\rho)$ never exceeds approximately 0.60995 bits.

The interpretation of this result is the following.  Even when one
has maximal information about a quantum system---i.e., one has a
pure state for it---one can predict almost nothing about the outcome
of a typical measurement \,\cite{Caves96}. In the limit of large $d$,
the outcome entropy for a typical measurement is just a little over a
half bit away from its maximal value.  Having a mixed state for a
system, reduces one's predictability even further, but indeed not by
that much: The small deviation is captured by the function in
Eq.~(\ref{Mittelstaedt}), which becomes a quantification of
uncertainty in its own right.

The way to get at a quantum statement of Eq.~(\ref{HungryBelly}) is
to make use of the fact that $S(\rho)$ and $Q(\rho)$ are both
concave in the variable $\rho$.\,\cite{Fuchs00b}  That is, for
either function, we have
\be
F\big(t\tilde\rho_0+(1-t)\tilde\rho_1\big)\ge t F(\tilde\rho_0)
+(1-t)F(\tilde\rho_1)\;,
\label{BingoBopp}
\ee
for any density operators $\tilde\rho_0$ and $\tilde\rho_1$ and any
real number $t\in[0,1]$.  Therefore, one might hope that
\be
F(\rho) \ge \sum_d P(d) F(\rho_d)\;.
\ee
Such a result however---if it is true---cannot arise in the trivial
fashion it did for the classical case of Eq.~(\ref{HungryBelly}).
This is because generally (as already emphasized),
\be
\rho\ne \sum_d P(d) \rho_d
\ee
for $\rho_d$ defined as in Eq.~(\ref{SlopFart}).  One therefore must
be more roundabout if a proof is going to happen.

The key is in noticing that
\bea
\rho &=& \rho^{1/2} I \rho^{1/2}
\nonumber
\\
&=& \sum_d \rho^{1/2}E_d \rho^{1/2}
\nonumber
\\
&=& \sum_d P(d) \tilde\rho_d
\label{HomericSimpsonic}
\eea
where
\be
\tilde\rho_d = \frac{1}{P(d)}\, \rho^{1/2}E_d \rho^{1/2} =
\frac{1}{P(d)}\, \rho^{1/2} A_d^\dagger A_d \rho^{1/2}\;.
\ee
What is special about this decomposition of $\rho$ is that for each
$d$, $\rho_d$ and $\tilde\rho_d$ have the same eigenvalues.  This
follows since $X^\dagger X$ and $X X^\dagger$ have the same
eigenvalues, for any operator $X$.  In the present case, setting
$X=A_d \rho^{1/2}$ does the trick.  Using the fact that both
$S(\rho)$ and $Q(\rho)$ depend only upon the eigenvalues of $\rho$
we obtain:
\bea
S(\rho) &\ge& \sum_d P(d) S(\rho_d)
\\
Q(\rho) &\ge& \sum_d P(d) Q(\rho_d)\;,
\eea
as we had been hoping for.  Thus, in performing an efficient quantum
measurement of a POVM $\{E_d\}$, an observer can expect to be left
with less uncertainty than he started with.\footnote{By differing
methods, a strengthening of this result in terms of a majorization
property can be found in Refs.~\cite{Fuchs00b} and
\cite{Nielsen00b}.}

In this sense, quantum ``collapse'' does indeed have some of the
flavor of Bayes' rule.  But we can expect more, and the derivation
above hints at just the right ingredient: $\rho_d$ and
$\tilde\rho_d$ have the same eigenvalues!  To see the impact of
this, let us once again explore the content of Eqs.~(\ref{SlopFart})
and (\ref{GrayJerseyDay}).  A common way to describe their meaning is
to use the operator polar-decomposition theorem\,\cite{Schatten60}
to rewrite Eq.~(\ref{SlopFart}) in the form
\be
\rho_d = \frac{1}{P(d)}\, U_d E_d^{1/2} \rho E_d^{1/2} U_d^\dagger\;,
\ee
where $U_d$ is a unitary operator. Since---subject only to the
constraint of efficiency---the operators $A_d$ are not determined any
further than Eq.~(\ref{GrayJerseyDay}), $U_d$ can be {\it any\/}
unitary operator whatsoever.  Thus, a customary way of thinking of
the state-change process is to break it up into two conceptual
pieces. First there is a ``raw collapse'':
\be
\rho\longrightarrow \sigma_d=\frac{1}{P(d)}\, E_d^{1/2} \rho
E_d^{1/2}\;.
\ee
Then, subject to the details of the measurement interaction and the
particular outcome $d$, one imagines the measuring device enforcing a
further kind of ``back-action'' or ``feedback'' on the measured
system\,\cite{FeedbackBoys}:
\be
\sigma_d\longrightarrow\rho_d=U_d\sigma_d U_d^\dagger\;.
\label{Dogma}
\ee
But this breakdown of the transition is a purely conceptual game.

Since the $U_d$ are arbitrary to begin with, we might as well break
down the state-change process into the following (nonstandard)
conceptual components. First one imagines an observer refining his
initial state of belief and simply plucking out a term corresponding
to the ``data'' collected:
\bea
\rho &=& \sum_d P(d) \tilde\rho_d
\label{RhythmStick}
\\
&& \phantom{\sum_d P(d)}\! \downarrow \nonumber
\\
\rho &&\stackrel{d}{\longrightarrow} \phantom{P(d)}\!\!\!
\tilde\rho_d\;.
\label{CoolEveningAir}
\eea
Finally, there may be a further ``mental readjustment'' of the
observer's beliefs, which takes into account details both of the
measurement interaction and the observer's initial quantum state.
This is enacted via some (formal) unitary operation $V_d$:
\be
\tilde\rho_d \longrightarrow \rho_d = V_d \tilde\rho_d V_d^\dagger\;.
\label{HitMe}
\ee
Putting the two processes together, one has the same result as the
usual picture.

The resemblance between the process in Eq.~(\ref{CoolEveningAir}) and
the classical Bayes' rule of Eq.~(\ref{Lasagna}) is
uncanny.\footnote{Other similarities between quantum collapse and
Bayesian conditionalization have been discussed in
Refs.~\cite{Bub77,Braunstein90,OzawaLump}.} By this way of viewing
things, quantum collapse is indeed not such a violent state of
affairs after all. Quantum measurement is nothing more, and nothing
less, than a refinement and a readjustment of one's initial state of
belief. More general state changes of the form Eq.~(\ref{StinkFart})
come about similarly, but with a further step of coarse-graining
(i.e., throwing away information that was in principle accessible).

Let us look at two limiting cases of efficient measurements.  In the
first, we imagine an observer whose initial belief structure
$\rho=|\psi\rangle\langle\psi|$ is a maximally tight state of belief.
By this account, no measurement whatsoever can {\it refine\/} it.
This follows because, no matter what $\{E_d\}$ is,
\be
\rho^{1/2}E_d \rho^{1/2} = P(d)|\psi\rangle\langle\psi|\;.
\ee
The only state change that can come about from a measurement must be
purely of the mental-readjustment sort:  We learn nothing new; we
just change what we can predict as a consequence of the side effects
of our experimental intervention.  That is to say, there is a sense
in which the measurement is solely disturbance. In particular, when
the POVM is an orthogonal set of projectors $\{\Pi_i=|i\rangle\langle
i|\}$ and the state-change mechanism is the von Neumann collapse
postulate, this simply corresponds to a readjustment according to
the unitary operators
\be
U_i=|i\rangle\langle\psi|\;.
\ee

At the opposite end of things, we can contemplate measurements that
have no possibility at all of causing a physical disturbance to the
system being measured. This could come about, for instance, by
interacting with one side of an entangled pair of systems and using
the consequence of that intervention to update one's beliefs about
the other side. In such a case, one can show that the state change is
purely of the refinement variety (with no further mental
readjustment).\footnote{This should be contrasted with the usual
picture of a ``minimally disturbing'' measurement of some POVM. In
our case, a minimal disturbance version of a POVM $\{E_d\}$
corresponds to taking $V_d=I$ for all $d$ in Eq.~(\ref{HitMe}).  In
the usual presentation---see Refs.~\cite{Fuchs00b} and
\cite{FeedbackBoys}---it corresponds to taking $U_d=I$ for all $d$ in
Eq.~(\ref{Dogma}) instead.  For instance, Howard Wiseman writes in
Ref.~\cite{FeedbackBoys}:
\bq
The action of $\big[ E_d^{1/2}\big]$ produces the minimum change in
the system, required by Heisenberg's relation, to be consistent with
a measurement giving the information about the state specified by the
probabilities [Eq.~(\ref{ChickenAndGravy})]. The action of [$U_d$]
represents additional back-action, an unnecessary perturbation of
the system. \ldots\ A back-action evading measurement is reasonably
defined by the requirement that, for all [$d$], [$U_d$] equals unity
(up to a phase factor that can be ignored without loss of
generality).
\eq
This of course means that, from the present point of view, there is
no such thing as a state-independent notion of minimally disturbing
measurement. Given an initial state $\rho$ and a POVM $\{E_d\}$, the
minimally disturbing measurement interaction is the one that
produces pure Bayesian updating with no further (purely quantum)
readjustment.} For instance, consider a pure state
$|\psi^{\scriptscriptstyle AB}\rangle$ whose Schmidt decomposition
takes the form
\be
|\psi^{\scriptscriptstyle AB}\rangle = \sum_i \sqrt{\lambda_i}
|a_i\rangle|b_i\rangle\;.
\ee
An efficient measurement on the $A$ side of this leads to a state
update of the form
\be
|\psi^{\scriptscriptstyle AB}\rangle\langle \psi^{\scriptscriptstyle
AB}|\, \longrightarrow\, (A_d\otimes I) |\psi^{\scriptscriptstyle
AB}\rangle\langle \psi^{\scriptscriptstyle AB}| (A_d^\dagger\otimes
I)\;.
\ee
Tracing out the $A$ side, then gives
\bea
\tr_{\scriptscriptstyle\rm A} \Big(A_d\otimes I
|\psi^{\scriptscriptstyle AB}\rangle
\langle \psi^{\scriptscriptstyle AB}| A_d^\dagger\otimes I\Big)
&=&
\sum_{ijk}\sqrt{\lambda_j}\sqrt{\lambda_k}\langle a_i|A_d\otimes I
|a_j\rangle|b_j\rangle\langle a_k|\langle b_k|A_d^\dagger\otimes
I|a_i\rangle
\nonumber
\\
&=&
\sum_{ijk}\sqrt{\lambda_j}\sqrt{\lambda_k}\langle a_k|A_d^\dagger
|a_i\rangle\langle a_i|A_d|a_j\rangle |b_j\rangle\langle b_k|
\nonumber
\\
&=&
\sum_{jk}\sqrt{\lambda_j}\sqrt{\lambda_k}\langle a_k|A_d^\dagger
A_d|a_j\rangle |b_j\rangle\langle b_k|
\nonumber
\\
&=&
\sum_{jk}\sqrt{\lambda_j}\sqrt{\lambda_k}\langle b_k|UA_d^\dagger
A_dU^\dagger|b_j\rangle |b_j\rangle\langle b_k|
\nonumber
\\
&=&
\sum_{jk}\sqrt{\lambda_j}\sqrt{\lambda_k}\langle
b_j|\Big(UA_d^\dagger A_dU^\dagger\Big)^{\!\scriptscriptstyle\rm
T}|b_k\rangle |b_j\rangle\langle b_k|
\nonumber
\\
&=&
\rho^{1/2} \Big(UA_d^\dagger
A_dU^\dagger\Big)^{\!\scriptscriptstyle\rm T} \rho^{1/2}
\eea
where $\rho$ is the initial quantum state on the $B$ side, $U$ is
the unitary operator connecting the $|a_i\rangle$ basis to the
$|b_i\rangle$ basis, and $^{\scriptscriptstyle\rm T}$ represents
taking a transpose with respect to the $|b_i\rangle$ basis.  Since
the operators
\be
F_d=\Big(UA_d^\dagger A_dU^\dagger\Big)^{\!\scriptscriptstyle\rm T}
\ee
go together to form a POVM, we indeed have the claimed result.

In summary, the lesson here is that it turns out to be rather easy to
think of quantum collapse as a noncommutative variant of Bayes'
rule.  In fact it is just in this that one starts to get a feel for
a further reason for Gleason's noncontextuality assumption. In the
setting of classical Bayesian conditionalization we have just that:
The probability of the transition $P(h)\longrightarrow P(h|d)$ is
governed solely by the local probability $P(d)$. The transition does
not care about how we have partitioned the rest of the potential
transitions.  That is, it does not care whether $d$ is embedded in a
two outcome set $\{d,\neg d\}$ or whether it is embedded in a three
outcome set, $\{d,e,\neg (d\vee e)\}$, etc. Similarly with the
quantum case. The probability for a transition from $\rho$ to
$\rho_0$ cares not whether our refinement is of the form
\be
\rho = P(0)\rho_0 + \sum_{d=1}^{17} P(d)\rho_d\quad \qquad \mbox{or
of the form}\quad \qquad \rho = P(0)\rho_0 + P(18)\rho_{18}\;,
\ee
as long as
\be
P(18)\rho_{18} = \sum_{d=1}^{17} P(d)\rho_d
\ee
What could be a simpler generalization of Bayes' rule?

Indeed, leaning on that, we can restate the discussion of the
``measurement problem'' at the beginning of Section 4 in slightly
more technical terms.  Go back to the classical setting of
Eqs.~(\ref{Orecchiette}) and (\ref{Macaroni}) where an agent has a
probability distribution $P(h,d)$ over two sets of hypotheses.
Marginalizing over the possibilities for $d$, one obtains the agent's
initial belief $P(h)$ about the hypothesis $h$. If he gathers an
explicit piece of data $d$, he should use Bayes' rule to update his
probability about $h$ to $P(h|d)$.

The question is this: Is the transition
\be
P(h) \longrightarrow P(h|d)
\ee
a mystery we should contend with?  If someone asked for a {\it
physical description\/} of this transition, would we be able to give
an explanation?  After all, one value for $h$ is true and always
remains true: there is no transition in it. One value for $d$ is
true and always remains true:  there is no transition in it. The
only discontinuous transition is in the {\it belief\/} $P(h)$, and
that presumably is a property of the believer's brain. To put the
issue into terms that start to sound like the quantum measurement
problem, let us ask: Should we not have a detailed theory of how the
brain works before we can trust in the validity of Bayes'
rule?\,\footnote{This point was recently stated much more eloquently
by Rocco Duvenhage in his paper Ref.~\cite{Duvenhage02}:
\bq
\indent In classical mechanics a measurement is nothing strange. It is
merely an event where the observer obtains information about some
physical system. A measurement therefore changes the observer's
information regarding the system. One can then ask: What does the
change in the observer's information mean? What causes it? And so
on. These questions correspond to the questions above, but now they
seem tautological rather than mysterious, since our intuitive idea
of information tells us that the change in the observer's
information simply means that he has received new information, and
that the change is caused by the reception of the new information.
We will see that the quantum case is no different \ldots

Let's say an observer has information regarding the state of a
classical system, but not necessarily complete information (this is
the typical case, since precise measurements are not possible in
practice). Now the observer performs a measurement on the system to
obtain new information \ldots The observer's information after this
measurement then differs from his information before the
measurement. In other words, a measurement ``disturbs'' the
observer's information. \ldots

The Heisenberg cut. This refers to an imaginary dividing line
between the observer and the system being observed \ldots It can be
seen as the place where information crosses from the system to the
observer, but it leads to the question of where exactly it should
be; where does the observer begin? In practice it's not really a
problem: It doesn't matter where the cut is. It is merely a
philosophical question which is already present in classical
mechanics, since in the classical case information also passes from
the system to the observer and one could again ask where the
observer begins. The Heisenberg cut is therefore no more problematic
in quantum mechanics than in classical mechanics.
\eq}

The answer is, ``Of course not!''  Bayes' rule---and beyond it all of
probability theory---is a tool that stands above the details of
physics. George Boole called probability theory a {\it law of
thought\/}\,\cite{Boole58}.  Its calculus specifies the optimal way
an agent should reason and make decisions when faced with incomplete
information.  In this way, probability theory is a generalization of
Aristotelian logic\footnote{In addition to
Ref.~\cite{JaynesPosthumous}, many further materials concerning this
point of view can be downloaded from the {\sl Probability Theory As
Extended Logic\/} web site maintained by G.~L. Bretthorst, {\tt
http://bayes.wustl.edu/}.}---a tool of thought few would accept as
being anchored to the details of the physical world.\footnote{We
have, after all, used simple Aristotelian logic in making deductions
from {\it all\/} our physical theories to date: from Aristotle's
physics to quantum mechanics to general relativity and even string
theory.} As far as Bayesian probability theory is concerned, a
``classical measurement'' is simply any {\it I-know-not-what\/} that
induces an application of Bayes' rule.  It is not the task of
probability theory (nor is it solvable within probability theory) to
explain how the transition Bayes' rule signifies comes about within
the mind of the agent.

The formal similarities between Bayes' rule and quantum collapse may
be telling us how to finally cut the Gordian knot of the measurement
problem. Namely, it may be telling us that it is simply not a problem
at all! Indeed, drawing on the analogies between the two theories,
one is left with a spark of insight:  perhaps the better part of
quantum mechanics is simply ``law of thought''\,\cite{Fuchs01a}.
Perhaps the structure of the theory denotes the optimal way to
reason and make decisions in light of {\it some\/} fundamental
situation---a fundamental situation waiting to be ferreted out in a
more satisfactory fashion.

This much we know:  That fundamental situation---whatever it
is---must be an ingredient Bayesian probability theory does not
have. As already emphasized, there must be something to drive a wedge
between the two theories. Probability theory alone is too general a
structure. Narrowing the structure will require input from the world
around us.

\subsection{Accepting Quantum Mechanics}

Looking at the issue from this perspective, let us ask: What does it
mean to accept quantum mechanics? Does it mean accepting (in
essence) the existence of an ``expert'' whose probabilities we
should strive to possess whenever we strive to be maximally
rational?\,\cite{vanFraassen02} The key to answering this question
comes from combining the previous discussion of Bayes' rule with the
considerations of the standard quantum-measurement device of Section
4.2.  For, contemplating this will allow us to go even further than
calling quantum collapse a noncommutative {\it variant\/} of Bayes'
rule.

Consider the description of quantum collapse in
Eqs.~(\ref{RhythmStick}) through (\ref{HitMe}) in terms of one's
subjective judgments for the outcomes of a standard quantum
measurement $\{E_h\}$.  Using the notation there, one starts with an
initial judgment
\be
P(h)=\tr(\rho E_h)
\ee
and, after a measurement of some other observable $\{E_d\}$, ends up
with a final judgment
\be
P_d(h)= \tr(\rho_d E_h)= \tr\big(\tilde\rho_d V_d^\dagger E_h
V_d\big) =\tr\big(\tilde\rho_d F^d_h\big)\;,
\ee
where
\be
F^d_h= V_d^\dagger E_h V_d\;.
\ee
Note that, in general, $\{E_h\}$ and $\{E_d\}$ refer to two entirely
different POVMs; the range of their indices $h$ and $d$ need not
even be the same.  Also, since $\{E_h\}$ is a minimal informationally
complete POVM, $\{F^d_h\}$ will itself be informationally complete
for each value of $d$.

Thus, {\it modulo a final unitary readjustment or redefinition of
the standard quantum measurement} based on the data gathered, one has
precisely Bayes' rule in this transition. This follows since
\be
\rho= \sum_d P(d)\tilde\rho_d
\ee
implies
\be
P(h)=\sum_d P(d) P(h|d)\;,
\ee
where
\be
P(h|d)=\tr(\tilde\rho_d E_h)\;.
\ee

Another way of looking at this transition is from the ``active''
point of view, i.e., that the axes of the probability simplex are
held fixed, while the state is transformed from $P(h|d)$ to
$P_d(h)$.  That is, writing
\be
F^d_h = \sum_{h'=1}^{D^2} \Gamma^d_{h h'} E_{h'}
\ee
where $\Gamma^d_{h h'}$ are some real-valued coefficients and
$\{E_{h'}\}$ refers to a relabeling of the original standard quantum
measurement, we get
\be
P_d(h)= \sum_{h'=1}^{D^2} \Gamma^d_{h h'} P(h'|d)\;.
\ee

This gives an enticingly simple description of what quantum
measurement {\it is\/} in Bayesian terms.  Modulo the final
readjustment, a quantum measurement is {\it any application of
Bayes' rule whatsoever\/} on the initial state $P(h)$.  By any
application of Bayes' rule, I mean in particular any convex
decomposition of $P(h)$ into some refinements $P(h|d)$ that also
live in ${\cal P}_{\rm\scriptscriptstyle SQM}$.\footnote{Note a
distinction between this way of posing Bayes' rule and the usual
way.  In stating it, I give no status to a joint probability
distribution $P(h,d)$.  If one insists on calling the product
$P(d)P(h|d)$ a joint distribution $P(h,d)$, one can do so of course,
but it is only a mathematical artifice without intrinsic meaning. In
particular, one should not get a feeling from $P(h,d)$'s
mathematical existence that the random variables $h$ and $d$
simultaneously coexist.  As always, $h$ and $d$ stand only for the
consequences of experimental interventions into nature; without the
intervention, there is no $h$ and no $d$.} Aside from the final
readjustment, a quantum measurement is just like a classical
measurement:  It is {\it any I-know-not-what\/} that pushes an agent
to an application of Bayes' rule.\footnote{Of course, I fear the
wrath my choice of words ``any I-know-not-what'' will bring down
upon me. For it will be claimed---I can see it now, rather
violently---that I do not understand the first thing of what the
``problem'' of quantum measurement is: It is to supply a mechanism
for understanding how collapse comes about, not to dismiss it. But
my language is honest language and meant explicitly to leave nothing
hidden.  The point here, as already emphasized in the classical
case, is that it is not the task---and cannot be the task---of a
theory that makes intrinsic use of probability to justify how an
agent has gotten hold of a piece of information that causes him to
change his beliefs.  A belief is a property of one's head, not of
the object of one's interest.}

\begin{figure} 
\begin{center}
\includegraphics[height=2in]{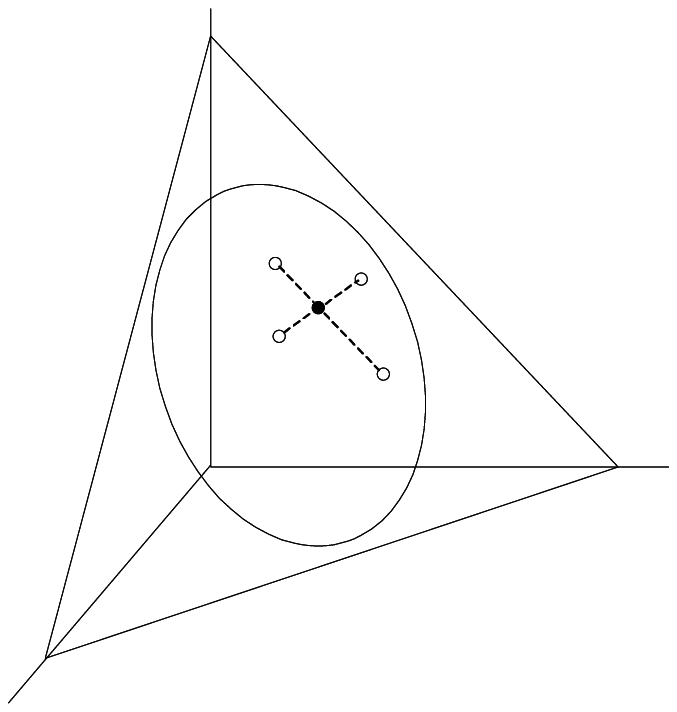}\hspace{1in}
\includegraphics[height=2in]{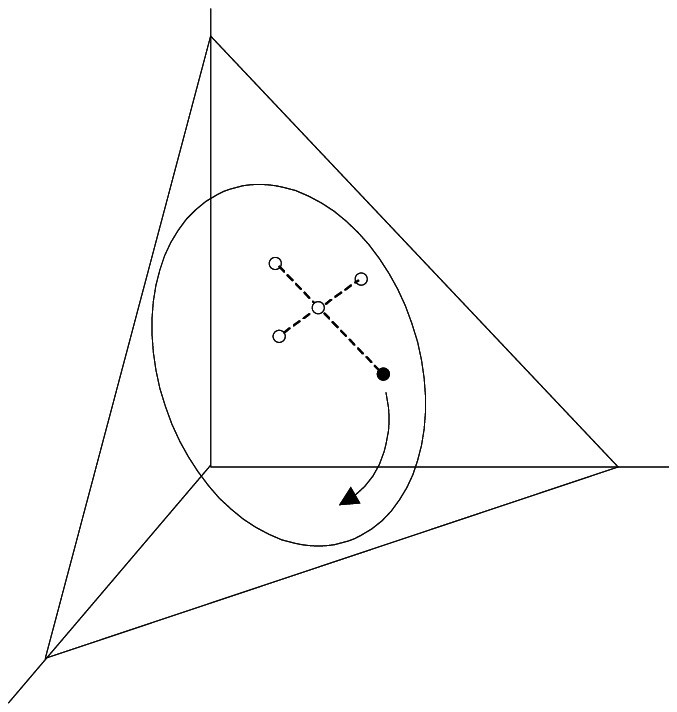}
\bigskip\caption{A quantum measurement is any ``I-know-not-what'' that
generates an application of Bayes' rule to one's beliefs for the
outcomes of a standard quantum measurement---that is, a
decomposition of the initial state into a convex combination of
other states and then a final ``choice'' (decided by the world, not
the observer) within that set. Taking into account the idea that
quantum measurements are ``invasive'' or ``disturbing'' alters the
classical Bayesian picture only in introducing a further
outcome-dependent readjustment:  On can either think of it passively
as a readjustment of the standard quantum measurement device, or
actively (as depicted here) as a further adjustment to the posterior
state.}
\end{center}
\end{figure}

Accepting the formal structure of quantum mechanics is---in large
part---simply accepting that it would not be in one's best interest
to hold a $P(h)$ that falls outside the convex set ${\cal
P}_{\rm\scriptscriptstyle SQM}$.  Moreover, up to the final
conditionalization rule signified by a unitary operator $V_d$, a
measurement is simply anything that can cause an application of
Bayes' rule within ${\cal P}_{\rm\scriptscriptstyle SQM}$.

But if there is nothing more than arbitrary applications of Bayes'
rule to ground the concept of quantum measurement, would not the
solidity of quantum theory melt away?  What else can determine when
``this'' rather than ``that'' measurement is performed?  Surely that
much has to be objective about the theory?

\section{What Else Is Information?}

\begin{flushright}
\parbox{3.8in}{\footnotesize
\bq
That's territory I'm not yet ready to follow you into. Good luck!
\\
\hspace*{\fill} --- {\it N. David Mermin}, 2002
\eq
}
\end{flushright}

Suppose one wants to hold adamantly to the idea that the quantum
state is purely subjective.  That is, that there is no right and true
quantum state for a system---the quantum state is ``numerically
additional'' to the quantum system.  It walks through the door when
the agent who is interested in the system walks through the door.
Can one consistently uphold this point of view at the same time as
supposing that {\it which\/} POVM $\{E_d\}$ and {\it which\/}
state-change rule $\rho\longrightarrow\rho_d=A_d\rho A_d^\dagger$ a
measurement device performs are objective features of the device?
The answer is no, and it is not difficult to see why.

Take as an example, a device that supposedly performs a standard von
Neumann measurement $\{\Pi_d\}$, the measurement of which is
accompanied by the standard collapse postulate.  Then when a click
$d$ is found, the posterior quantum state will be $\rho_d=\Pi_d$
regardless of the initial state $\rho$.  If this state-change rule
is an objective feature of the device or its interaction with the
system---i.e., it has nothing to do with the observer's subjective
judgment---then the final state $\rho_d$ too must be an objective
feature of the quantum system. The argument is that simple.
Furthermore, it clearly generalizes to all state change rules for
which the $A_d$ are rank-one operators without adding any further
complications.

Also though, since the operators $E_d$ control the maximal
support\,\footnote{The support of an operator is the subspace
spanned by its eigenvectors with nonzero eigenvalues\@.} of the
final state $\rho_d$ through $A_d=U_d E_d^{1/2}$, it must be that
even the $E_d$ themselves are subjective judgments. For otherwise,
one would have a statement like, ``Only states with support within a
subspace ${\cal S}_d$ are {\it correct}.  All other states are
simply {\it wrong}.''\footnote{Such a statement, in fact, is not so
dissimilar to the one found in Ref.~\cite{Brun02}.  For several
rebuttals of that idea, see Ref.~\cite{Fuchs02a} and
\cite{CavesFuchsSchack02}.}

Thinking now of uninterrupted quantum time evolution as the special
case of what happens to a state after the single-element POVM
$\{I\}$ is performed, one is forced to the same conclusion even in
that case.  The time evolution super-operator for a quantum
system---most generally a completely positive trace-preserving
linear map on the space of operators for ${\cal
H}_{\scriptscriptstyle\rm D}$\,\cite{Kraus83}---is a subjective
judgment on exactly the same par as the subjectivity of the quantum
state.

Here is another way of seeing the same thing.  Recall what I viewed
to be the most powerful argument for the quantum state's
subjectivity---the Einsteinian argument of Section 3.  Since we can
toggle the quantum state from a distance, it must not be something
sitting over there, but rather something sitting over here:  It can
only be our information about the far-away system. Let us now apply
a variation of this argument to time evolutions.

\begin{figure} 
\begin{center}
\includegraphics[height=2in]{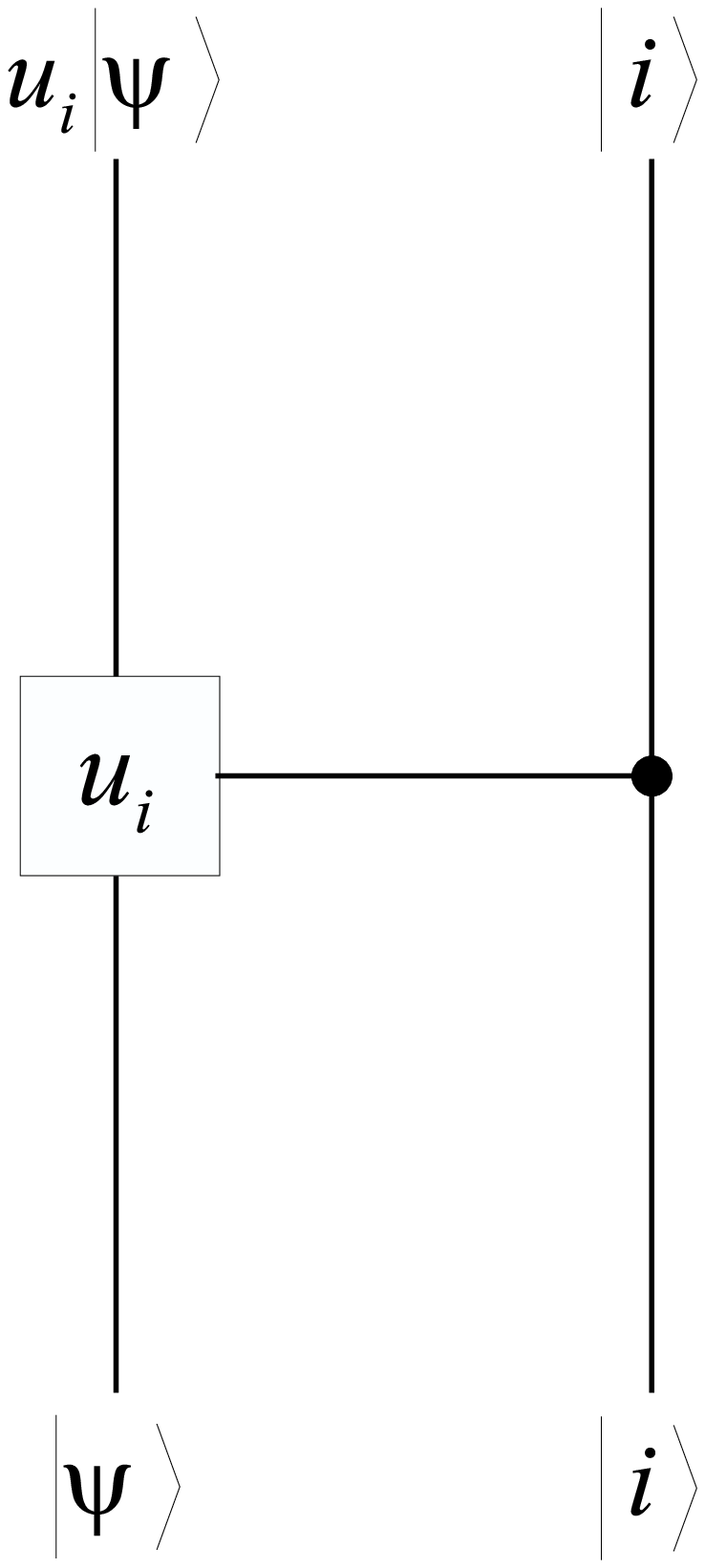}\hspace{1.2in}
\includegraphics[height=2in]{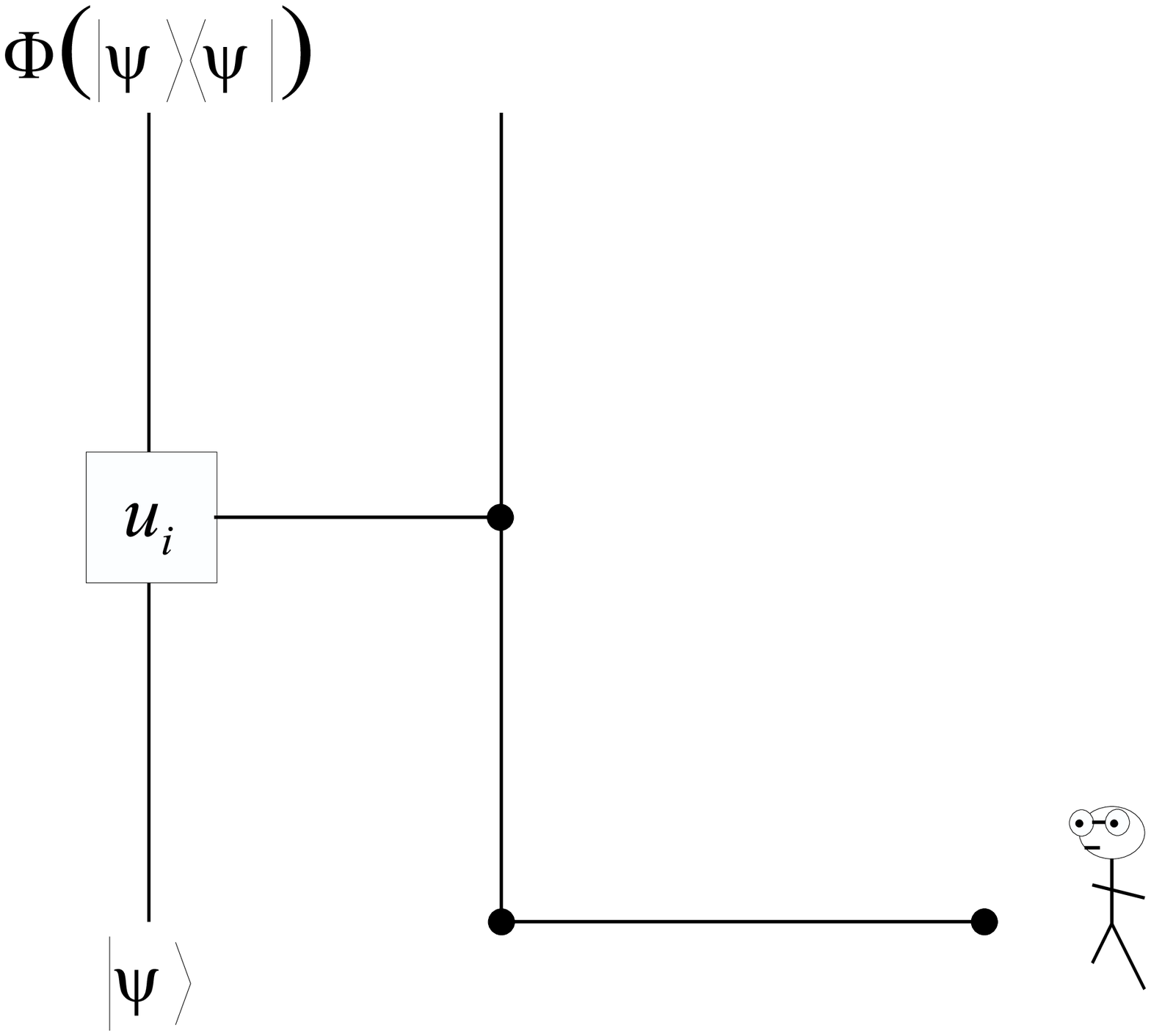}
\bigskip\caption{One can use a slight modification of Einstein's
argument for the subjectivity of the quantum state to draw the same
conclusion for quantum time evolutions.  By performing measurements
on a far away system, one will ascribe one or another completely
positive map to the evolution of the left-most qubit.  Therefore,
accepting physical locality, the time evolution map so ascribed
cannot be a property intrinsic to the system.}
\end{center}
\end{figure}

Consider a simple quantum circuit on a bipartite quantum system that
performs a controlled unitary operation $U_i$ on the target bit.
(For simplicity, let us say the bipartite system consists of two
qubits.)  Which unitary operation the circuit applies depends upon
which state $|i\rangle$, $i=0,1$, of two orthogonal states impinges
upon the control bit. Thus, for an arbitrary state $|\psi\rangle$ on
the target, one finds
\be
|i\rangle|\psi\rangle\;\longrightarrow\; |i\rangle \big(U_i
|\psi\rangle\big)
\ee
for the overall evolution.  Consequently the evolution of the target
system alone is given by
\be
|\psi\rangle\;\longrightarrow\; U_i |\psi\rangle
\ee
On the other hand, suppose the control bit is prepared in a
superposition state $|\phi\rangle = \alpha|0\rangle +
\beta|1\rangle$.  Then the evolution for the target bit will be given
by a completely positive map $\Phi_\phi$.  That is,
\be
|\psi\rangle\;\longrightarrow\; \Phi_\phi(|\psi\rangle\langle\psi|)=
|\alpha|^2\, U_0|\psi\rangle\langle\psi| U_0^\dagger +|\beta|^2\,
U_1|\psi\rangle\langle\psi| U_1^\dagger\;.
\ee

Now, to the point.  Suppose rather than feeding a single qubit into
the control bit, we feed half of an entangled pair, where the other
qubit is physically far removed from the circuit.  If an observer
with this description of the whole set-up happens to make a
measurement on the far-away qubit, then he will be able to induce any
of a number of completely positive maps $\Phi_\phi$ on the control
bit.  These will depend upon which measurement he performs and which
outcome he gets.  The point is the same as before:  Invoking
physical locality, one obtains that the time evolution mapping on
the single qubit cannot be an objective state of affairs localized
at that qubit.  The time evolution, like the state, is subjective
information.\,\footnote{Of course, there are sideways moves one can
use to try to get around this conclusion.  For instance, one could
argue that, ``The time evolution operator $\Phi$ on the control
qubit is only an `effective' evolution for it.  The `true' evolution
for the system is the unitary evolution specified by the complete
quantum circuit.''\cite{Caves99}  In my opinion, however, moves like
this are just prostrations to the Everettic temple.  One could
dismiss the original Einsteinian argument in the same way: ``The
observer toggles nothing with his localized measurement; the `true'
quantum state is the universal quantum state.  All that is going on
in a quantum measurement is the revelation of a relative
state---i.e., the `effective' quantum state.'' How can one argue with
this, other than to say it is not the most productive stance and that
the evidence shows that since 1957 it has not been able to quell the
foundations debate.  See Footnote~{\ref{GodandAllah}}.}\footnote{A
strengthening of this argument may also be found in the same way as
in Section 3:  Namely, by considering the teleportation of quantum
dynamics.  I will for the moment, however, leave that as an exercise
for the reader. See the many references in Ref.~\cite{VidalLump} for
appropriate background.}

It has long been known that the trace preserving completely positive
linear maps $\Phi$ over a $D$-dimensional vector space can be placed
in a one-to-one correspondence with density operators on a
$D^2$-dimensional space via the
relation\,\cite{Schumacher96,Jamiolkowski72,Choi75}
\be
\Upsilon=I\otimes\Phi\big(|\psi_{\rm\scriptscriptstyle
ME}\rangle\langle \psi_{\rm\scriptscriptstyle ME}|\big)
\ee
where $|\psi_{\rm\scriptscriptstyle ME}\rangle$ signifies a
maximally entangled state on ${\cal H}_{\rm\scriptscriptstyle
D}\otimes {\cal H}_{\rm\scriptscriptstyle D}$,
\be
|\psi_{\rm\scriptscriptstyle
ME}\rangle=\frac{1}{\sqrt{D}}\sum_{i=1}^D |i\rangle|i\rangle\;.
\ee
This is usually treated as a convenient representation theorem only,
but maybe it is no mathematical accident. Perhaps there is a deep
physical reason for it: The time evolution one ascribes to a quantum
system IS a density operator! It is a quantum state of belief no
more and no less than the initial quantum state one assigns to that
same system.

How to think about this?  Let us go back to the issue that closed
the last section.  How can one possibly identify the meaning of a
measurement in the Bayesian view, where a measurement ascription is
itself subjective---i.e., a measurement finds a mathematical
expression only in the subjective refinement of some agent's beliefs?
Here is the difficulty.  When one agent contemplates viewing a piece
of data $d$, he might be willing to use the data to refine his
beliefs according to
\be
P(h)=\sum_d P(d)P(h|d)\;.
\ee
However there is nothing to stop another agent from thinking the
same data warrants him to refine his beliefs according to
\be
Q(h)=\sum_d Q(d)Q(h|d)\;.
\ee
A priori, there need be no relation between the $P$'s and the
$Q$'s.

A relation only comes when one seeks a criterion for when the two
agents will say that they believe they are drawing the same meaning
from the data they obtain. That identification is a purely voluntary
act; for there is no way for the agent to walk outside of his
beliefs and see the world as it completely and totally is. The
standard Bayesian solution to the problem is this: When both agents
accept the same ``statistical model'' for their expectations of the
data $d$ given a hypothesis $h$, then they will agree to the
identity of the measurements they are each (separately) considering.
I.e., two agents will say they are performing the same measurement
when and only when
\be
P(d|h)=Q(d|h)\;,\quad \forall h \mbox{ and } \forall d\;.
\ee
Putting this in a more evocative form, we can say that both agents
agree to the meaning of a measurement when they adopt the same
resolution of the identity
\be
1=\sum_d \frac{P(d)P(h|d)}{P(h)}=\sum_d \frac{Q(d)Q(h|d)}{Q(h)}\;.
\ee
with which to describe it.

With this, the relation to quantum measurement should be apparent. If
we take it seriously that a measurement is anything that generates a
refinement of one's beliefs, then an agent specifies a measurement
when he specifies a resolution of his initial density operator
\be
\rho=\sum_d P(d)\,\tilde\rho_d\;.
\ee
But again, there is nothing to stop another agent from thinking the
data warrants a refinement that is completely unrelated to the first:
\be
\sigma=\sum_d Q(d)\,\tilde\sigma_d\;.
\ee
And that is where the issue ends if the agents have no further
agreement.

Just as in the classical case, however, there is a solution for the
identification problem. Using the canonical construction of
Eq.~(\ref{HomericSimpsonic}), we can say that both agents agree to
the meaning of a measurement when they adopt the same resolution of
the identity,
\be
I\;=\;\sum_d P(d)\,\rho^{-1/2}\tilde\rho_d\,\rho^{-1/2}\;=\; \sum_d
Q(d)\,\sigma^{-1/2}\tilde\sigma_d\,\sigma^{-1/2}
\ee
with which to describe it.

Saying it in a more tautological way, two agents will be in agreement
on the identity of a measurement when they assign it the same POVM
$\{E_d\}$,
\be
E_d \;=\; P(d)\,\rho^{-1/2}\tilde\rho_d\,\rho^{-1/2}\;=\;
Q(d)\,\sigma^{-1/2}\tilde\sigma_d\,\sigma^{-1/2}\;.
\ee
The importance of this move, however, is that it draws out the
proper way to think about the operators $E_d$ from the present
perspective.  They play part of the role of the ``statistical model''
$P(d|h)$.  More generally, that role is fulfilled by the complete
state change rule:
\be
P(d|h) \qquad \longleftrightarrow \qquad \rho\rightarrow\rho_d
\ee
That is to say, drawing the correspondence in different terms,
\be
P(d|h) \qquad \longleftrightarrow \qquad \Phi_d(\cdot)=U_d
E_d^{1/2}\cdot E_d^{1/2} U_d^\dagger\;.
\ee
(Of course, more generally---for nonefficient
measurements---$\Phi_d(\cdot)$ may consist of a convex sum of such
terms.)

The completely positive map that gives a mathematical description to
quantum time evolution is just such a map.  Its role is that of the
subjective statistical model $P(d|h)$, where $d$ just happens to be
drawn from a one-element set.

Thus, thinking back on entanglement, it seems the general structure
of quantum time evolutions cannot the wedge we are looking for
either. What we see instead is that there is a secret waiting to be
unlocked, and when it is unlocked, it will very likely tell us as
much about quantum time evolutions as quantum states and quantum
measurements.

\section{Intermission}

Let us take a deep breath.  Up until now I have tried to trash about
as much quantum mechanics as I could, and I know that takes a
toll---it has taken one on me.  Section 3 argued that quantum
states---whatever they are---cannot be objective entities.  Section 4
argued that there is nothing sacred about the quantum probability
rule and that the best way to think of a quantum state is as a state
of belief about what {\it would\/} happen if one were to ever
approach a standard measurement device locked away in a vault in
Paris. Section 5 argued that even our hallowed quantum entanglement
is a secondary and subjective effect. Section 6 argued that all a
measurement {\it is\/} is just an arbitrary application of Bayes'
rule---an arbitrary refinement of one's beliefs---along with some
account that measurements are invasive interventions into nature.
Section 7 argued that even quantum time evolutions are subjective
judgments; they just so happen to be conditional judgments.  \ldots\
And, so it went.

Subjective.  Subjective!  Subjective!!  It is a word that will not go
away.  But subjectivity is not something to be worshipped for its
own sake. There are limits: The last thing we need is a bloodbath of
deconstruction. At the end of the day, there had better be some
term, some element in quantum theory that stands for the objective,
or we might as well melt away and call this all a dream.

I turn now to a more constructive phase.

\section{Unknown Quantum States?}
\begin{flushright}
\parbox{4.0in}{\footnotesize
\bq
My thesis, paradoxically, and a little pro\-vocatively, but
nonetheless genuinely, is simply this:
\begin{center}
QUANTUM STATES DO NOT EXIST.
\end{center}
The abandonment of superstitious beliefs about the existence of
Phlogiston, the Cosmic Ether, Absolute Space and Time, ..., or
Fairies and Witches, was an essential step along the road to
scientific thinking. The quantum state, too, if regarded as
something endowed with some kind of objective existence, is no less
a misleading conception, an illusory attempt to exteriorize or
materialize our true probabilistic beliefs.\smallskip
\\
\hspace*{\fill} ---~{\it the~true~ghost~of~Bruno~de~Finetti}
\eq
}
\end{flushright}

The hint of a more fruitful direction can be found by trying to make
sense of one of the most commonly used phrases in quantum information
theory from a Bayesian perspective. It is the {\it unknown quantum
state}. There is hardly a paper in quantum information that does not
make use of it. Unknown quantum states are
teleported\,\cite{Bennett93}, protected with quantum error
correcting codes\,\cite{Shor95}, and used to check for quantum
eavesdropping\,\cite{Bennett84}.  The list of uses grows each day.
But what can the term mean?  In an information-based interpretation
of quantum mechanics, it is an oxymoron: If quantum states, by their
very definition, are states of subjective information and not states
of nature, then the state is {\it known\/} by someone---at the very
least, by the person who wrote it down.

Thus, if a phenomenon ostensibly invokes the concept of an unknown
state in its formulation, that unknown state had better be shorthand
for a more basic situation (even if that basic situation still awaits
a complete analysis).  This means that for any phenomenon using the
idea of an unknown quantum state in its description, we should
demand that either
\begin{enumerate}
\item
The owner of the unknown state---a further decision-making agent or
observer---be explicitly identified.  (In this case, the unknown
state is merely a stand-in for the unknown {\it state of belief\/} of
an essential player who went unrecognized in the original
formulation.)  Or,
\item
If there is clearly no further agent or observer on the scene, then
a way must be found to reexpress the phenomenon with the term
``unknown state'' completely banished from its formulation. (In this
case, the end-product of the effort will be a single quantum state
used for describing the phenomenon---namely, the state that actually
captures the describer's overall set of beliefs throughout.)
\end{enumerate}

This Section reports the work of Ref.~\cite{Caves01} and
\,\cite{Schack00}, where such a project is carried out for the
experimental practice of {\it quantum-state
tomography\/}\,\cite{Vogel89}. The usual description of tomography is
this.  A device of some sort, say a nonlinear optical medium driven
by a laser, repeatedly prepares many instances of a quantum system,
say many temporally distinct modes of the electromagnetic field, in
a fixed quantum state $\rho$, pure or mixed\,\cite{vanEnk01}.  An
experimentalist who wishes to characterize the operation of the
device or to calibrate it for future use may be able to perform
measurements on the systems it prepares even if he cannot get at the
device itself. This can be useful if the experimenter has some prior
knowledge of the device's operation that can be translated into a
probability distribution over states. Then learning about the state
will also be learning about the device. Most importantly, though,
this description of tomography assumes that the precise state $\rho$
is unknown.  The goal of the experimenter is to perform enough
measurements, and enough kinds of measurements (on a large enough
sample), to estimate the identity of $\rho$.

This is clearly an example where there is no further player on whom
to pin the unknown state as a state of belief or judgment.  Any
attempt to find such a missing player would be entirely artificial:
Where would the player be placed?  On the inside of the device the
tomographer is trying to characterize?\footnote{Placing the player
here would be about as respectable as George Berkeley's famous patch
to his philosophical system of idealism. The difficulty is captured
engagingly by a limerick of Ronald Knox and its anonymous reply:
\bq
There was a young man who said, ``God {\bf :} Must think it
exceedingly odd {\bf :} If he finds that this tree {\bf :} Continues
to be {\bf :} When there's no one about in the Quad.'' REPLY: ``Dear
Sir: Your astonishment's odd. {\bf :} I am always about in the Quad.
{\bf :} And that's why the tree {\bf :} Will continue to be, {\bf :}
Since observed by Yours faithfully, God.''\eq} The only available
course is the second strategy above---to banish the idea of the
unknown state from the formulation of tomography.

To do this, we once again take our cue from Bayesian probability
theory\,\cite{Kyburg80,JaynesPosthumous,Bernardo94}. As emphasized
previously, in Bayesian theory probabilities---just like quantum
states---are not objective states of nature, but rather measures of
belief, reflecting one's operational commitments in various gambling
scenarios. In light of this, it comes as no surprise that one of the
most overarching Bayesian themes is to identify the conditions under
which a set of decision-making agents can come to a common belief or
probability assignment for a random variable even though their
initial beliefs may differ\,\cite{Bernardo94}. Following that theme
is the key to understanding the essence of quantum-state tomography.

Indeed, classical Bayesian theory encounters almost precisely the
same problem as our unknown quantum state through the widespread use
of the phrase ``unknown probability'' in its domain.  This is an
oxymoron every bit as egregious as unknown state.

The procedure analogous to quantum-state tomography in Bayesian
theory is the estimation of an unknown probability from the results
of repeated trials on ``identically prepared systems.'' The way to
eliminate unknown probabilities from this situation was introduced
by Bruno de Finetti in the early 1930s \,\cite{DeFinetti90}.  His
method was simply to focus on the equivalence of the repeated
trials---namely, that what is really important is that the systems
are indistinguishable as far as probabilistic predictions are
concerned.  Because of this, any probability assignment
$p(x_1,x_2,\ldots,x_N)$ for multiple trials should be symmetric
under permutation of the systems.  As innocent as this conceptual
shift may sound, de Finetti was able to use it to powerful effect.
For, with his {\it representation theorem}, he showed that any
multi-trial probability assignment that is permutation-symmetric for
an arbitrarily large number of trials---de Finetti called such
multi-trial probabilities {\it exchangeable\/}---is equivalent to a
probability for the ``unknown probabilities.''

Let us outline this in a little more detail. In an objectivist
description of $N$ ``identically prepared systems,'' the individual
trials are described by discrete random variables
$x_n\in\{1,2,\ldots,k\}$, $n=1,\ldots,N$, and the probability in the
multi-trial hypothesis space is given by an independent identically
distributed distribution
\begin{equation}
p(x_1,x_2,\ldots,x_N)\,=\,p_{x_1} p_{x_2} \cdots p_{x_N}\, =\,
p_1^{n_{\scriptscriptstyle 1}} p_2^{n_{\scriptscriptstyle 2}}\cdots
p_k^{n_{\scriptscriptstyle k}}\;.
\label{eq-iid}
\end{equation}
The numbers $p_j$ describe the objective, ``true'' probability that
the result of a single experiment will be $j$ ($j=1,\ldots,k$).  The
variable $n_j$, on the other hand, describes the number of times
outcome $j$ is listed in the vector $(x_1,x_2,\ldots,x_N)$. But this
description---for the objectivist---only describes the situation from
a kind of ``God's eye'' point of view.  To the experimentalist, the
``true'' probabilities $p_1,\ldots,p_k$ will very often be {\it
unknown\/} at the outset.  Thus, his burden is to estimate the
unknown probabilities by a statistical analysis of the experiment's
outcomes.

In the Bayesian approach, however, it does not make sense to talk
about estimating a true probability. Instead, a Bayesian assigns a
prior probability distribution $p(x_1,x_2,\ldots,x_N)$ on the
multi-trial hypothesis space and uses Bayes' theorem to update the
distribution in the light of his measurement results. The content of
de Finetti's theorem is this. Assuming {\it only\/} that
\be
p(x_{\pi(1)},x_{\pi(2)},\ldots,x_{\pi(N)}) = p(x_1,x_2,\ldots,x_N)
\ee
for any permutation $\pi$ of the set $\{1,\ldots,N\}$, and that for
any integer $M>0$, there is a distribution
$p_{N+M}(x_1,x_2,\ldots,x_{N+M})$ with the same permutation property
such that
\be
p(x_1,x_2,\ldots,x_N)\; =
\sum_{x_{N+1},\ldots,x_{N+M}}p_{N+M}(x_1,\ldots,x_N,x_{N+1},\ldots,x_{N+M})
\;,
\label{eq-marginal}
\ee
then $p(x_1,x_2,\ldots,x_N)$ can be written uniquely in the form
\bea
p(x_1,x_2,\ldots,x_N)&=& \int_{{\cal S}_k} P(\vec{p}\,)\,p_{x_1}
p_{x_2} \cdots p_{x_N}\,d\vec{p}
\label{Hestia}
\nonumber
\\
&=&
\int_{{\cal S}_k} P(\vec{p}\,)\, p_1^{n_{\scriptscriptstyle 1}}
p_2^{n_{\scriptscriptstyle 2}}\cdots p_k^{n_{\scriptscriptstyle k}}
\, d\vec{p}\;,
\label{eq-repr}
\eea
where $\vec{p}=(p_1,p_2,\ldots,p_k)$, and the integral is taken over
the simplex of such distributions
\be
{\cal S}_k=\left\{\vec{p}\mbox{ : }\; p_j\ge0\mbox{ for all } j\mbox{
and } \sum_{j=1}^k p_j=1\right\}.
\ee
Furthermore, the function $P(\vec{p}\,)\ge0$ is required to be a
probability density function on the simplex:
\begin{equation}
\int_{{\cal S}_k} P(\vec{p}\,)\,d\vec{p}=1\;,
\end{equation}
With this representation theorem, the unsatisfactory concept of an
unknown probability vanishes from the description in favor of the
fundamental idea of assigning an exchangeable probability
distribution to multiple trials.

With this cue in hand, it is easy to see how to reword the
description of quantum-state tomography to meet our goals.  What is
relevant is simply a judgment on the part of the
experimenter---notice the essential subjective character of this
``judgment''---that there is no distinction between the systems the
device is preparing.  In operational terms, this is the judgment
that {\it all the systems are and will be the same as far as
observational predictions are concerned}.  At first glance this
statement might seem to be contentless, but the important point is
this: To make this statement, one need never use the notion of an
unknown state---a completely operational description is good enough.
Putting it into technical terms, the statement is that if the
experimenter judges a collection of $N$ of the device's outputs to
have an overall quantum state $\rho^{(N)}$, he will also judge any
permutation of those outputs to have the same quantum state
$\rho^{(N)}$. Moreover, he will do this no matter how large the
number $N$ is. This, complemented only by the consistency condition
that for any $N$ the state $\rho^{(N)}$ be derivable from
$\rho^{(N+1)}$, makes for the complete story.

The words ``quantum state'' appear in this formulation, just as in
the original formulation of tomography, but there is no longer any
mention  of {\it unknown\/} quantum states.  The state $\rho^{(N)}$
is known by the experimenter (if no one else), for it represents his
judgment.  More importantly, the experimenter is in a position to
make an unambiguous statement about the structure of the whole
sequence of states $\rho^{(N)}$: Each of the states $\rho^{(N)}$ has
a kind of permutation invariance over its factors. The content of
the {\it quantum de Finetti representation
theorem}\,\cite{Caves01,Hudson76} is that a sequence of states
$\rho^{(N)}$ can have these properties, which are said to make it an
{\it exchangeable\/} sequence, if and only if each term in it can
also be written in the form
\begin{equation}
\rho^{(N)}=\int_{{\cal D}_{\rm\scriptscriptstyle D}} P(\rho)\,
\rho^{\otimes N}\, d\rho\;,
\label{Jeremy}
\end{equation}
where $\rho^{\otimes N}=\rho\otimes\rho\otimes\cdots\otimes\rho$ is
an $N$-fold tensor product. Here $P(\rho)\ge0$ is a fixed probability
distribution over the density operator space ${\cal
D}_{\rm\scriptscriptstyle D}$, and
\begin{equation}
\int_{{\cal D}_{\rm\scriptscriptstyle D}} P(\rho)\,d\rho=1\;,
\end{equation}
where $d\rho$ is a suitable measure.

The interpretive import of this theorem is paramount. For it alone
gives a mandate to the term unknown state in the usual description of
tomography.  It says that the experimenter can act {\it as if\/} his
judgment $\rho^{(N)}$ comes about because he knows there is a ``man
in the box,'' hidden from view, repeatedly preparing the same state
$\rho$.  He does not know which such state, and the best he can say
about the unknown state is captured in the probability distribution
$P(\rho)$.

The quantum de Finetti theorem furthermore makes a connection to the
overarching theme of Bayesianism stressed above.  It guarantees for
two independent observers---as long as they have a rather minimal
agreement in their initial beliefs---that the outcomes of a
sufficiently informative set of measurements will force a
convergence in their state assignments for the remaining
systems\,\cite{Schack00}.  This ``minimal'' agreement is
characterized by a judgment on the part of both parties that the
sequence of systems is exchangeable, as described above, and a
promise that the observers are not absolutely inflexible in their
opinions. Quantitatively, the latter means that though $P(\rho)$ may
be arbitrarily close to zero, it can never vanish.

This coming to agreement works because an exchangeable density
operator sequence can be updated to reflect information gathered
from measurements by another quantum version of Bayes's rule for
updating probabilities\,\cite{Schack00}. Specifically, if
measurements on $K$ systems yield results $D_K$, then the state of
additional systems is constructed as in Eq.~(\ref{Jeremy}), but
using an updated probability on density operators given by
\begin{equation}
P(\rho|D_K)={P(D_K|\rho)P(\rho)\over P(D_K)}\;.
\label{QBayes}
\end{equation}
Here $P(D_K|\rho)$ is the probability to obtain the measurement
results $D_K$, given the state $\rho^{\otimes K}$ for the $K$
measured systems, and
\be
P(D_K)=\int_{{\cal D}_{\rm\scriptscriptstyle D}}
P(D_K|\rho)\,P(\rho)\,d\rho
\ee
is the unconditional probability for the measurement results. For a
sufficiently informative set of measurements, as $K$ becomes large,
the updated probability $P(\rho|D_K)$ becomes highly peaked on a
particular state $\rho_{D_K}$ dictated by the measurement results,
regardless of the prior probability $P(\rho)$, as long as $P(\rho)$
is nonzero in a neighborhood of $\rho_{D_K}$. Suppose the two
observers have different initial beliefs, encapsulated in different
priors $P_i(\rho)$, $i=1,2$.  The measurement results force them to a
common state of belief in which any number $N$ of additional systems
are assigned the product state $\rho_{D_K}^{\otimes N}$, i.e.,
\begin{equation}
\int P_i(\rho|D_K)\,\rho^{\otimes N}\,d\rho
\quad{\longrightarrow}\quad \rho_{D_K}^{\otimes N}\;,
\label{HannibalLecter}
\end{equation}
independent of $i$, for $K$ sufficiently large.

This shifts the perspective on the purpose of quantum-state
tomography:  It is not about uncovering some ``unknown state of
nature,'' but rather about the various observers' coming to
agreement over future probabilistic predictions. In this connection,
it is interesting to note that the quantum de Finetti theorem and
the conclusions just drawn from it work only within the framework of
complex vector-space quantum mechanics. For quantum mechanics based
on real Hilbert spaces\,\cite{Stueckelberg60}, the connection between
exchangeable density operators and unknown quantum states does not
hold.

A simple counterexample is the following.  Consider the $N$-system
state
\begin{equation}
\rho^{(N)}={1\over2}\rho_+^{\otimes N} + {1\over2}\rho_-^{\otimes N}
\;,
\label{eq-real}
\end{equation}
where
\begin{equation}
\rho_+={1\over2}(I+\sigma_2)\qquad\mbox{and}
\qquad\rho_-={1\over2}(I-\sigma_2)
\end{equation}
and $\sigma_1$, $\sigma_2$, and $\sigma_3$ are the Pauli matrices.
In complex-Hilbert-space quantum mechanics, Eq.~(\ref{eq-real}) is
clearly a valid density operator:  It corresponds to an equally
weighted mixture of $N$ spin-up particles and $N$ spin-down
particles in the $y$-direction.  The state $\rho^{(N)}$ is thus
exchangeable, and the decomposition in Eq.~(\ref{eq-real}) is unique
according to the quantum de Finetti theorem.

But now consider $\rho^{(N)}$ as an operator in real-Hilbert-space
quantum mechanics.  Despite its ostensible use of the imaginary
number $i$, it remains a valid quantum state.  This is because, upon
expanding the right-hand side of Eq.~(\ref{eq-real}), all the terms
with an odd number of $\sigma_2$'s cancel away.  Yet, even though it
is an exchangeable density operator, it cannot be written in de
Finetti form Eq.~(\ref{Jeremy}) using only real symmetric operators.
This follows because $i\sigma_2$ cannot be written as a linear
combination of $I$, $\sigma_1$, and $\sigma_3$, while a
real-Hilbert-space de Finetti expansion as in Eq.~(\ref{Jeremy}) can
{\it only\/} contain those three operators. Hence the de Finetti
theorem does not hold in real-Hilbert-space quantum mechanics.

In classical probability theory, exchangeability characterizes those
situations where the only data relevant for updating a probability
distribution are frequency data, i.e., the numbers $n_j$ in
Eq.~(\ref{eq-repr}). The quantum de Finetti representation shows
that the same is true in quantum mechanics: Frequency data (with
respect to a sufficiently robust measurement, in particular, any one
that is informationally complete) are sufficient for updating an
exchangeable state to the point where nothing more can be learned
from sequential measurements. That is, one obtains a convergence of
the form Eq.~(\ref{HannibalLecter}), so that ultimately any further
measurements on the individual systems will be statistically
independent. That there is no quantum de Finetti theorem in real
Hilbert space means that there are fundamental differences between
real and complex Hilbert spaces with respect to learning from
measurement results.

Finally, in summary, let us hang on the point of learning for just a
little longer.  The quantum de Finetti theorem shows that the
essence of quantum-state tomography is not in revealing an ``element
of reality'' but in deriving that various agents (who agree some
minimal amount) can come to agreement in their ultimate quantum-state
assignments. This is not at all the same thing as the statement
``reality does not exist.''  It is simply that one need not go to
the extreme of taking the ``unknown quantum state'' as being
objectively real to make sense of the experimental practice of
tomography.

J.~M. Bernardo and A.~F.~M. Smith in their book
Ref.~\cite{Bernardo94} word the goal of these exercises we have run
through in this paper very nicely:
\bq
\indent
[I]ndividual degrees of belief, expressed as probabilities, are
inescapably the starting point for descriptions of uncertainty.
There can be no theories without theoreticians; no learning without
learners; in general, no science without scientists.  It follows
that learning processes, whatever their particular concerns and
fashions at any given point in time, are necessarily reasoning
processes which take place in the minds of individuals.  To be sure,
the object of attention and interest may well be an assumed
external, objective reality:  but the actuality of the learning
process consists in the evolution of individual, subjective beliefs
about that reality. However, it is important to emphasize \ldots\
that the primitive and fundamental notions of {\it individual\/}
preference and belief will typically provide the starting point for
{\it interpersonal\/} communication and reporting processes. \ldots\
[W]e shall therefore often be concerned to identify and examine
features of the individual learning process which relate to
interpersonal issues, such as the conditions under which an
approximate consensus of beliefs might occur in a population of
individuals.
\eq
The quantum de Finetti theorem provides a case in point for how much
agreement a population can come to from within quantum mechanics.

One is left with a feeling---an almost salty feeling---that perhaps
this is the whole point of the structure of quantum mechanics.
Perhaps the missing ingredient for narrowing the structure of
Bayesian probability down to quantum mechanics has been in front of
us all along. It finds no better expression than in taking account
of the challenges the physical world poses to our coming to
agreement.

\section{The Oyster and the Quantum}
\begin{flushright}
\parbox{4.0in}{\footnotesize
\bq
\indent The significance of this development is to give us insight into
the logical possibility of a new and wider pattern of thought.  This
takes into account the observer, including the apparatus used by him,
differently from the way it was done in classical physics \ldots\ In
the new pattern of thought we do not assume any longer the {\it
detached observer}, occurring in the idealizations of this classical
type of theory, but an observer who by his indeterminable effects
creates a new situation, theoretically described as a new state of
the observed system. \ldots In this way every observation is a
singling out of a particular factual result, here and now, from the
theoretical possibilities, thereby making obvious the discontinuous
aspect of the physical phenomena.

Nevertheless, there remains still in the new kind of theory an {\it
objective reality}, inasmuch as these theories deny any possibility
for the observer to influence the results of a measurement, once the
experimental arrangement is chosen. Particular qualities of an
individual observer do not enter the conceptual framework of the
theory.
\\
\hspace*{\fill} --- {\it Wolfgang Pauli}, 1954
\eq
}
\end{flushright}

A grain of sand falls into the shell of an oyster and the result is
a pearl.  The oyster's sensitivity to the touch is the source a
beautiful gem. In the 75 years that have passed since the founding
of quantum mechanics, only the last 10 have turned to a view and an
attitude that may finally reveal the essence of the theory.  The
quantum world is sensitive to the touch, and that may be one of the
best things about it.  Quantum information---with its three
specializations of quantum information theory, quantum cryptography,
and quantum computing---leads the way in telling us how to quantify
that idea.  Quantum algorithms can be exponentially faster than
classical algorithms.  Secret keys can be encoded into physical
systems in such a way as to reveal whether information has been
gathered about them.  The list of triumphs keeps growing.

The key to so much of this has been simply in a change of attitude.
This can be seen by going back to almost any older textbook on
quantum mechanics: Nine times out of ten, the Heisenberg uncertainty
relation is presented in a way that conveys the feeling that we have
been short-changed by the physical world.

\bq
\small
``Look at classical physics, how nice it is:  We can measure a
particle's position and momentum with as much accuracy as we would
like.  How limiting quantum theory is instead.  We are stuck with
$$
\Delta x \Delta p \ge \frac{1}{2}\hbar\;,
$$
and there is nothing we can do about it. The task of physics is to
sober up to this state of affairs and make the best of it.''
\eq
How this contrasts with the point of departure of quantum
information!  There the task is not to ask what limits quantum
mechanics places upon us, but what novel, productive things we can
do in the quantum world that we could not have done otherwise. In
what ways is the quantum world fantastically better than the
classical one?

If one is looking for something ``real'' in quantum theory, what more
direct tack could one take than to look to its technologies? People
may argue about the objective reality of the wave function ad
infinitum, but few would argue about the existence of quantum
cryptography as a solid prediction of the theory. Why not take that
or a similar effect as the grounding for what quantum mechanics is
trying to tell us about nature?

Let us try to give this imprecise set of thoughts some shape by
reexpressing quantum cryptography in the language built up in the
previous sections.  For quantum key distribution it is essential to
be able to prepare a physical system in one or another quantum state
drawn from some fixed {\it nonorthogonal\/} set
\,\cite{Bennett84,Bennett92}. These nonorthogonal states are used to
encode a potentially secret cryptographic key to be shared between
the sender and receiver. The information an eavesdropper seeks is
about which quantum state was actually prepared in each individual
transmission. What is novel here is that the encoding of the
proposed key into nonorthogonal states forces the
information-gathering process to induce a disturbance to the overall
{\it set\/} of states. That is, the presence of an active
eavesdropper transforms the initial pure states into a set of mixed
states or, at the very least, into a set of pure states with larger
overlaps than before. This action ultimately boils down to a loss of
predictability for the sender over the outcomes of the receiver's
measurements and, so, is directly detectable by the receiver (who
reveals some of those outcomes for the sender's inspection). More
importantly, there is a direct connection between the statistical
information gained by an eavesdropper and the consequent disturbance
she must induce to the quantum states in the process.  As the
information gathered goes up, the necessary disturbance also goes up
in a precisely formalizable way\,\cite{Fuchs96,SpekkensLump}.

Note the two ingredients that appear in this scenario. First, the
information gathering or measurement is grounded with respect to one
observer (in this case, the eavesdropper), while the disturbance is
grounded with respect to another (here, the sender). In particular,
the disturbance is a disturbance to the sender's previous
information---this is measured by her diminished ability to predict
the outcomes of certain measurements the legitimate receiver might
perform. No hint of any variable intrinsic to the system is made use
of in this formulation of the idea of ``measurement causing
disturbance.''

The second ingredient is that one must consider at least two possible
non\-orthogonal preparations in order for the formulation to have
any meaning. This is because the information gathering is not about
some classically-defined observable---i.e., about some unknown
hidden variable or reality intrinsic to the system---but is instead
about which of the unknown states the sender actually prepared.  The
lesson is this: Forget about the unknown preparation, and the random
outcome of the quantum measurement is information about nothing.  It
is simply ``quantum noise'' with no connection to any preexisting
variable.

How crucial is this second ingredient---that is, that there be at
least two nonorthogonal states within the set under consideration?
We can address its necessity by making a shift in the account above:
One might say that the eavesdropper's goal is not so much to uncover
the identity of the unknown quantum state, but to sharpen her
predictability over the receiver's measurement outcomes.  In fact,
she would like to do this at the same time as disturbing the
sender's predictions as little as possible.  Changing the language
still further to the terminology of Section 4, the eavesdropper's
actions serve to sharpen her information about the potential
consequences of the receiver's further interventions on the system.
(Again, she would like to do this while minimally diminishing the
sender's previous information about those same consequences.) In the
cryptographic context, a byproduct of this effort is that the
eavesdropper ultimately comes to a more sound prediction of the
secret key. From the present point of view, however, the importance
of this change of language is that it leads to an almost Bayesian
perspective on the information--disturbance problem.

As previously emphasized, within Bayesian probability the most
significant theme is to identify the conditions under which a set of
decision-making agents can come to a common probability assignment
for some random variable in spite of the fact that their initial
probabilities differ\,\cite{Bernardo94}.  One might similarly view
the process of quantum eavesdropping.  The sender and the
eavesdropper start off initially with differing quantum state
assignments for a single physical system.  In this case it so
happens that the sender can make sharper predictions than the
eavesdropper about the outcomes of the receiver's measurements.  The
eavesdropper, not satisfied with this situation, performs a
measurement on the system in an attempt to sharpen those
predictions.  In particular, there is an attempt to come into
something of an agreement with the sender but without revealing the
outcomes of her measurements or, indeed, her very presence.

It is at this point that a distinct {\it property\/} of the quantum
world makes itself known.  The eavesdropper's attempt to
surreptitiously come into alignment with the sender's predictability
is always shunted away from its goal.  This shunting of various
observer's predictability is the subtle manner in which the quantum
world is sensitive to our experimental interventions.

And maybe this is our crucial hint!  The wedge that drives a
distinction between Bayesian probability theory in general and
quantum mechanics in particular is perhaps nothing more than this
``Zing!''~of a quantum system that is manifested when an agent
interacts with it. It is this wild sensitivity to the touch that
keeps our information and beliefs from ever coming into too great of
an alignment. The most our beliefs about the potential consequences
of our interventions on a system can come into alignment is captured
by the mathematical structure of a pure quantum state $|\psi\rangle$.
Take all possible information-disturbance curves for a quantum
system, tie them into a bundle, and {\it that\/} is the long-awaited
property, the input we have been looking for from nature. Or, at
least, that is the speculation.

\subsection{Give Us a Little Reality}
\begin{flushright}
\parbox{2.8in}{\footnotesize
\bq
What we need here is a little Realit$t$y.
\\
\hspace*{\fill} --- {\it Herbert Bernstein}, circa 1997
\eq
}
\end{flushright}

In the previous version of this paper\,\cite{Fuchs01b} I ended the
discussion just at this point with the following words, ``Look at
that bundle long and hard and we might just find that it stays
together without the help of our tie.''  But I imagine that wispy
command was singularly unhelpful to anyone who wanted to pursue the
program further.

How might one hope to mathematize the bundle of all possible
information-disturbance curves for a system?  If it can be done at
all, the effort will have to end up depending upon a single real
parameter---the dimension of the system's Hilbert space. As a safety
check, let us ask ourselves right at the outset whether this is a
tenable possibility?  Or will Hilbert-space dimension go the wayside
of subjectivity, just as we saw so many of the other terms in the
theory go?  I think the answer will be in the negative: Hilbert-space
dimension will survive to be a stand-alone concept with no need of an
agent for its definition.

The simplest check perhaps is to pose the Einsteinian test for it as
we did first for the quantum state and then for quantum time
evolutions. Posit a bipartite system with Hilbert spaces ${\cal
H}_{\rm\scriptscriptstyle D_1}$ and ${\cal H}_{\rm\scriptscriptstyle
D_2}$ (with dimensions $D_1$ and $D_2$ respectively) and imagine an
initial quantum state for that bipartite system.  As argued too many
times already, the quantum state must be a subjective component in
the theory because the theory allows localized measurements on the
$D_1$ system to change the quantum state for the $D_2$ system. In
contrast, is there anything one can do at the $D_1$ site to change
the numerical value of $D_2$?  It does not appear so.  Indeed, the
only way to change that number is to scrap the initial supposition.
Thus, to that extent, one has every right to call the numbers $D_1$
and $D_2$ potential ``elements of reality.''

It may not look like much, but it is a start.\footnote{Cf.\ also
Ref.~\cite{Blume-Kohout02}.} And one should not belittle the power of
a good hint, no matter how small.\footnote{Cf.\ also the final
paragraphs of Section 1.}

\section{Appendix: Changes Made Since {\tt quant-ph/0106166} Version}

Beside overhauling the Introduction so as to make it more relevant to
the present meeting, I made the following more substantive changes
to the old version:

\begin{enumerate}
\itemsep -1pt

\item
I made the language slightly less flowery throughout.
\item
Some of the jokes are now explained for the readers who thought they
were typographical errors.
\item
For the purpose of Section 1's imagery, I labeled the followers of
the Spontaneous Collapse and Many-Worlds interpretations,
Spontaneous Collapseans and Everettics---in contrast to the previous
terms Spontaneous Collapsicans and Everettistas---to better emphasize
their religious aspects.
\item
Some figures were removed from the quantum de Finetti section and
the dramatis personae on page 2 was added.
\item
I now denote the outcomes of a general POVM by the index $d$ to
evoke the image that all (and only) a quantum measurement ever does
is gather a piece of {\it data\/} by which we update our subjective
probabilities for something else.  It causes us to change our
subjective probability assignments $P(h)$ for some hypothesis $h$ to
a posterior assignment $P_d(h)$ conditioned on the data $d$.
\item
As noted in Footnote \ref{obfuscatoid}, this paper is a bit of a
transitionary one for me in that, since writing {\tt
quant-ph/0106166}, I have become much more convinced of the
consistency and value of the ``radically'' subjective Bayesian
paradigm for probability theory. That is, I have become much more
inclined to the view of Bruno de Finetti\,\cite{DeFinetti90}, say,
than that of Edwin Jaynes\,\cite{JaynesApp}. To that end, I have
stopped calling probability distributions ``states of knowledge''
and been more true to the conception that they are ``states of
belief'' whose cash-value is determined by the way an agent will
gamble in light of them. That is, a probability distribution, once
it is written down, is very literally a gambling commitment the
writer of it uses with respect to the phenomenon he is describing.
It is not clear to what extent this adoption of terminology will
cause obfuscation rather than clarity in the present paper; it was
certainly not needed for many of the discussions.  Still I could not
stand to propagate my older view any further.
\item
In general, 23 footnotes, 38 equations, and over 43 references have
been added. There are five new historical quotes starting the
sections, and the ghostly quote of Section 9 has been modified for
greater accuracy.
\item
The metaphor ending Section 1, describing how the grail of the
present quantum foundations program can be likened to the spacetime
manifold of general relativity, is new.
\item
Section 2 has been expanded to be consistent with the rest of the
paper.  Also, there are three important explanatory footnotes to be
found there.
\item
Einstein's letter to Michele Besso in Section 3 is now quoted in
full.
\item
Section 4.1, which argues more strongly for Gleason's
noncontextuality assumption than previously, is new.
\item
Section 4.2, which explains informationally complete POVMs and uses
them to imagine a ``standard quantum measurement'' at the Bureau of
Weights and Measures, is new.
\item
To elaborate the connection between entanglement and the standard
probability rule, I switched the order of presentation of the
``Whither Bayes Rule?''~and ``Wither Entanglement?''~sections.
\item
The technical mistake that was in Section 5 is now deleted.  The
upshot of the old argument, however, remains:  The tensor-product
rule for combining quantum systems can be thought of as secondary to
the structure of local observables.
\item
A much greater elaboration of the ``classical measurement
problem''---i.e., the mystery of {\it physical\/} cause of Bayesian
conditionalization upon the acquisition of new information (or the
lack of a mystery thereof)---is now given in Section 6.
\item
Section 6.1, wherein a more detailed description of the relation
between real-world measurements and the hypothetical standard quantum
measurement is fleshed out, is new.
\item
Section 7, which argues for the nonreality of the Hamiltonian and
the necessary subjectivity of the ascription of a POVM to a
measurement device, is new.
\item
Section 8, wherein I find a way to use the word bloodbath, is new.
\item
The long quote in Section 9 by Bernardo and Smith, which describes what
Bayesian probability theory strives for, is new.  Here's another good
quote of theirs that didn't fit in anywhere else:
\bq
\indent
What is the nature and scope of Bayesian Statistics within this
spectrum of activity?

Bayesian Statistics offers a rationalist theory of personalistic
beliefs in contexts of uncertainty, with the central aim of
characterising how an individual should act in order to avoid certain
kinds of undesirable behavioural inconsistencies.  The theory
establishes that expected utility maximization provides the basis for
rational decision making and that Bayes' theorem provides the key to
the ways in which beliefs should fit together in the light of
changing evidence.  The goal, in effect, is to establish rules and
procedures for individuals concerned with disciplined uncertainty
accounting.  The theory is not descriptive, in the sense of claiming
to model actual behaviour.  Rather, it is prescriptive, in the sense
of saying ``if you wish to avoid the possibility of these undesirable
consequences you must act in the following way.''
\eq
\item
Section 10.1, which argues for the nonsubjectivity of Hilbert-space
dimension, is new.
\item
One can read about the term ``Realit$t$y'' in Ref.~\cite{Fortun98}.
\end{enumerate}

\section{Acknowledgments}

I thank Carl Caves, Greg Comer, David Mermin, and R\"udiger Schack
for the years of correspondence that led to this view, Jeff Bub and
Lucien Hardy for giving me courage in general, Ad\'an Cabello, Asher
Peres, and Arkady Plotnitsky for their help in compiling the
dramatis personae of the Introduction, Jeff Nicholson for composing
the paper's figures, and Andrei Khrennikov for infinite patience.
Further thanks go to Charlie Bennett, Matthew Donald, Steven van
Enk, Jerry Finkelstein, Philippe Grangier, Osamu Hirota, Andrew
Landahl, Hideo Mabuchi, Jim Malley, Mike Nielsen, Masanao Ozawa,
John Preskill, Terry Rudolph, Johann Summhammer, Chris Timpson, and
Alex Wilce for their many comments on the previous version of this
paper---all of which I tried to respond to in some shape or
fashion---and particularly warm gratitude goes to Howard Barnum for
pointing out my technical mistake in the ``Wither
Entanglement?''~section. Finally, I thank Ulrich Mohrhoff for
calling me a Kantian; it taught me that I should work a little
harder to make myself look Jamesian.

\end{document}